# Status Report of the DPHEP Collaboration: A Global Effort for Sustainable Data Preservation in High Energy Physics

www.dphep.org

## Abstract

Data from High Energy Physics (HEP) experiments are collected with significant financial and human effort and are mostly unique. An inter-experimental study group on HEP data preservation and long-term analysis was convened as a panel of the International Committee for Future Accelerators (ICFA). The group was formed by large collider-based experiments and investigated the technical and organisational aspects of HEP data preservation. An intermediate report was released in November 2009 addressing the general issues of data preservation in HEP and an extended blueprint paper was published[1] in 2012. In July 2014 the DPHEP collaboration was formed as a result of the signature of the Collaboration Agreement by seven large funding agencies (others have since joined or are in the process of acquisition) and in June 2015 the first DPHEP Collaboration Workshop[2] and Collaboration Board meeting took place.

This status report of the DPHEP collaboration details the progress during the period 2013 – 2015 inclusive.

International Collaboration for Data Preservation and Long Term Analysis in High Energy Physics

[1] See http://arxiv.org/pdf/1205.4667.
[2] See https://indico.cern.ch/event/377026/other-view?view=standard.

## Executive Summary

- Significant progress has been made in the past years regarding our understanding of, and implementation of services and solutions for, long-term data preservation for future re-use;

- **However, continued investment in data preservation is needed: without this the data will soon become unusable or indeed lost (as history has told us all too many times);**

- **Some of this investment can be done centrally, e.g. by providing bit preservation services for multiple experiments at a given laboratory, whilst important elements need to be addressed on an experiment-by-experiment basis.**

- Funding agencies – and indeed the general public – are now understanding the need for preservation and sharing of "data" (which typically includes significant metadata, software and "knowledge") with requirements on data management plans, preservation of data, reproducibility of results and sharing of data and results becoming increasingly important and in some cases mandatory;

- The "business case" for data preservation in scientific, educational and cultural as well as financial terms is increasingly well understood: funding beyond (or outside) the standard lifetime of projects is required to ensure this preservation;

- A well-established model for data preservation exists – the Open Archival Information System (OAIS). Whilst developed primarily in the Space Data Community, it has since been adopted by all most all disciplines – ranging from Science to Humanities and Digital Cultural Heritage – and provides useful terminology and guidance that has proven applicable also to HEP;

- **The main message – from Past and Present Circular Colliders to Future ones – is that it is never early to consider data preservation: early planning is likely to result in cost savings that may be significant. Furthermore, resources (and budget) beyond the data-taking lifetime of the projects must be foreseen from the beginning.**

# Introduction

Shortly after the publication of the DPHEP Blueprint (see below), various inputs concerning the long-term preservation of HEP data were made to the group preparing the update to the European Strategy for Particle Physics. An updated strategy was adopted by a special session[3] of the CERN Council in May 2013 in Brussels, and this[4] includes the following statement:

> The success of particle physics experiments, such as those required for the high-luminosity LHC, relies on innovative instrumentation, state-of-the-art infrastructures and large-scale data-intensive computing. *Detector R&D programmes should be supported strongly at CERN, national institutes, laboratories and universities. Infrastructure and engineering capabilities for the R&D programme and construction of large detectors, as well as infrastructures for data analysis, **data preservation** and distributed data-intensive computing **should be maintained and further developed.***

As of 2013, with the appointment by CERN of a DPHEP Project Manager – one of the priorities identified in the Blueprint – the first steps towards transitioning to a Collaboration began. Seven institutes signed the Collaboration Agreement of May 2014, with additional (and often active) partners preparing to join.

After numerous workshops organized by and involving the Study Group, topical workshops on the "Full Costs of Curation" (January 2014)[5] and on "Common Projects and Shared Use Cases" (June 2015)[6] have been held.

The former has been instrumental in ensuring medium to long-term funding for the data preservation resources needed by the LHC experiments, whereas several CERN groups have committed support and services needed for the primary Use Cases agreed by these experiments (see below), which in many cases is matched by effort from the experiments and/or external institutes.

> **The message that constant effort and investment is needed should not be lost. However this effort can be well justified by the measurable benefits. These include not only direct benefits to the (sometimes former) collaboration in terms of scientific papers and PhDs obtained, but also in terms of much needed publicity for HEP through educational outreach and "open access" activities.**

Future events where data preservation experiences and solutions can be shared will continue, as well as topical events as needs arise. (An event[7] is planned in conjunction with WLCG in Lisbon in February 2016, to prepared a detailed Data Preservation Plan following the OAIS and related standards.

---

[3] See https://indico.cern.ch/event/244974/page/1.
[4] See https://indico.cern.ch/event/244974/page/7.
[5] See https://indico.cern.ch/event/276820/.
[6] See https://indico.cern.ch/event/377026/.
[7] See http://indico.cern.ch/event/433164/.

## The DPHEP Study Group

The DPHEP study group was initiated in early 2009 and became a sub-group of the International Committee for Future Accelerators (ICFA) – emphasizing its global nature – later that year. Its goal was:

***High Energy Physics*** *experiments initiate with this* **Study Group**[8] *a common reflection on* ***data persistency and long-term analysis*** *in order to get a common vision on these issues and create a multi-experiment dynamics for further reference.*

***The objectives of the Study Group are:***

- *Review and document the physics objectives of the data persistency in HEP.*
- *Exchange information concerning the analysis model: abstraction, software, documentation etc. and identify coherence points.*
- *Address the hardware and software persistency status.*
- *Review possible funding programs and other related international initiatives.*
- *Converge to a common set of specifications in a document that will constitute the basis for future collaborations.*

As well as running a series of workshops that rotated around all of the main HEP laboratories, it generated a Blueprint document that was well received by ICFA and was fed into the process for updating the European Strategy for Particle Physics.

The full Blueprint – which runs close to 100 pages – should be referred to for details regarding the motivation for and status of data preservation activities across all key laboratories (status in 2012).

It states:

*"Data from high-energy physics (HEP) experiments are collected with significant financial and human effort and are mostly unique. An inter-experimental study group on HEP data preservation and long-term analysis was convened as a panel of the International Committee for Future Accelerators (ICFA). The group was formed by large collider-based experiments and investigated the technical and organisational aspects of HEP data preservation. An intermediate report was released in November 2009 addressing the general issues of data preservation in HEP. This paper includes and extends the intermediate report. It provides an analysis of the research case for data preservation and a detailed description of the various projects at experiment, laboratory and international levels. In addition, the paper provides a concrete proposal for an international organisation in charge of the data management and policies in high-energy physics."*

The DPHEP study group identified the following priorities, in order of urgency:

- ***Priority 1: Experiment Level Projects in Data Preservation.*** *Large laboratories should define and establish data preservation projects in order to avoid catastrophic loss of data once major collaborations come to an end. The recent expertise gained during the last three years indicate that an extension of the computing effort within experiments with a person-power of the order of 2-3 FTEs leads to a significant improvement in the ability to move to a long-term data preservation phase. Such initiatives exist already or are being defined in the participating laboratories and are followed attentively by the study group.*

[8] See http://dphep.org for further details.

- ***Priority 2: International Organisation DPHEP.*** *The efforts are best exploited by a common organisation at the international level. The installation of this body, to be based on the existing ICFA study group, requires a Project Manager (1 FTE) to be employed as soon as possible. The effort is a joint request of the study group and could be assumed by rotation among the participating laboratories.*

- ***Priority 3: Common R&D projects.*** *Common requirements on data preservation are likely to evolve into inter-experimental R&D projects (three concrete examples are given above, each involving 1-2 dedicated FTE, across several laboratories). The projects will optimise the development effort and have the potential to improve the degree of standardisation in HEP computing in the longer term. Concrete requests will be formulated in common by the experiments to the funding agencies and the activity of these projects will be steered by the DPHEP organisation.*

*These priorities could be enacted with a funding model implying synergies from the three regions (Europe, America, Asia) and strong connections with laboratories hosting the data samples.*

# The DPHEP Collaboration Agreement

In order to implement priority 2 above (experiment-level data preservation is already under way in most cases and common "R&D" projects are already leading to services with a view to long-term support and sustainability), CERN has appointed a Project Manager (October 2012) and a Collaboration Agreement has been prepared. 9 institutes have now signed this agreement (CERN, DESY, HIP Finland, IHEP, IN2P3, KEK, MPP, IPP and STFC[9]) with several more in the pipeline.

The agreement, which largely reflects the recommendations of the Blueprint, includes the following goals:

*The Project, in coordination with the International Committee for Future Accelerators (ICFA), aims at:*

1. *Positioning itself as the natural forum for the entire discipline in order to foster discussion, achieve consensus and transfer knowledge in two main areas:*

    a. *Technological challenges in data preservation in HEP,*
    b. *Diverse governance at the collaboration and community level for preserved data,*

2. *Co-ordinate common R&D projects aiming to establish common, discipline-wide preservation tools,*
3. *Harmonize preservation projects across the Partners and liaise with relevant initiatives from other fields,*
4. *Design the long-term organization of sustainable and economic preservation in HEP,*
5. *Outreach within the community and advocacy towards the main stakeholders for the case of preservation in HEP.*

All of these areas are currently being pursued actively and can be viewed in terms of a (slowly evolving) "2020 vision".

[9] Not yet formally ratified by a DPHEP Collaboration Board meeting.

## The DPHEP Collaboration and Implementation Boards

The DPHEP Collaboration Board (see Appendix A) consists of a representative of all of the institutes / bodies that have signed the DPHEP Collaboration Agreement. One meeting has been held so far, immediately after the DPHEP Collaboration Workshop of June 2015 and future meetings will be held approximately annually. The meetings are also open to future members of the Collaboration, as well as representatives from key projects such as DASPOS[10].

The DPHEP Implementation Board (see Appendix B) meets more regularly and is composed of active participants in data preservation for HEP. The meeting frequency has dropped somewhat with time, given the relative maturity of a number of the data preservation projects as well as the occurrence of focused meetings on specific topics / technologies, such as CernVM, analysis capture and preservation and so forth.

The agendas of the meetings, as well as any material presented, can be found via the Indico category: https://indico.cern.ch/category/4458/. A web archive of the corresponding mailing list (requires authentication) can be found at https://groups.cern.ch/group/DPHEP-IB/default.aspx.

## The DPHEP 2020 Vision

The "vision" for DPHEP – first presented to ICFA in February 2013 – a consists of the following key points:

- By 2020, all **archived data** – e.g. that described in DPHEP Blueprint, including LHC data – should be easily **findable** and fully **usable** by the **designated communities** with clear (Open) access policies and possibilities to annotate further

- Best practices, tools and services should be well run-in, fully documented and sustainable; built in common with **other disciplines**, based on standards

- There should be a **DPHEP portal**, through which data / tools accessed

- **Clear targets & metrics** to measure the above should be agreed between **Funding Agencies, Service Providers** and the **Experiments (Collaborations).**

Although there is clearly much work still to be done, this vision looks both achievable and the timescale for realizing it has been significantly reduced through interactions with other (non-HEP) projects and communities.

## Requirements from Funding Agencies

There have been numerous policy discussions and recommendations in recent years, some of which are reflected in the outputs of the (EU FP7) projects discussed below.

[10] Data and Software Preservation for Open Science – see https://daspos.crc.nd.edu/.

A particularly clear statement can be found from the US Office of Science[11] that includes the following:

***All proposals submitted to the Office of Science (after 1 October 2014) for research funding must include a Data Management Plan (DMP) that addresses the following requirements:***

- ***DMPs should describe whether and how data generated in the course of the proposed research will be shared and preserved.***

  *If the plan is not to share and/or preserve certain data, then the plan must explain the basis of the decision*

  ***At a minimum, DMPs must describe how data sharing and preservation will enable validation of results, or how results could be validated if data are not shared or preserved***

Similar requirements are coming (or have come) from other Funding Agencies and for International projects in particular it will be important to understand how to respond to these in a consistent manner. This is part of the debate that will continue, e.g. following the RECODE project recommendations covered below.

# Open Access Policies

The four main LHC experiments have approved Open Access policies[12] that, whilst they differ in detail, are broadly similar (and are being adopted by other experiments):

1. (Moving towards) Gold Open Access for Publications (DPHEP "level 1");
2. Open Access to Specific Data Samples for Outreach (DPHEP "level 2");
3. Open Access to (a fraction of the) Reconstructed data (after an embargo period) (DPHEP "level 3");
4. Raw data[13] closed even to collaboration (today) (DPHEP "level 4").

| **Preservation Model (DPHEP Level)** | **Use case** |
|---|---|
| 1. Provide additional documentation | Publication-related information search |
| 2. Preserve the data in a simplified format | Outreach, simple training analyses |
| 3. Preserve the analysis level software and data format | Full scientific analysis based on existing reconstruction |
| 4. Preserve the reconstruction and simulation software and basic level data | Full potential of the experimental data |

Table 1 - DPHEP Preservation Models (Levels) - from the DPHEP Blueprint 2012

[11] See http://science.energy.gov/funding-opportunities/digital-data-management/.
[12] See http://opendata.cern.ch/collection/data-policies.
[13] Most disciplines use a different notation, with "L0" corresponding to the raw data and L1/L2/L3 corresponding to calibrated and/or processed and/or derived data.

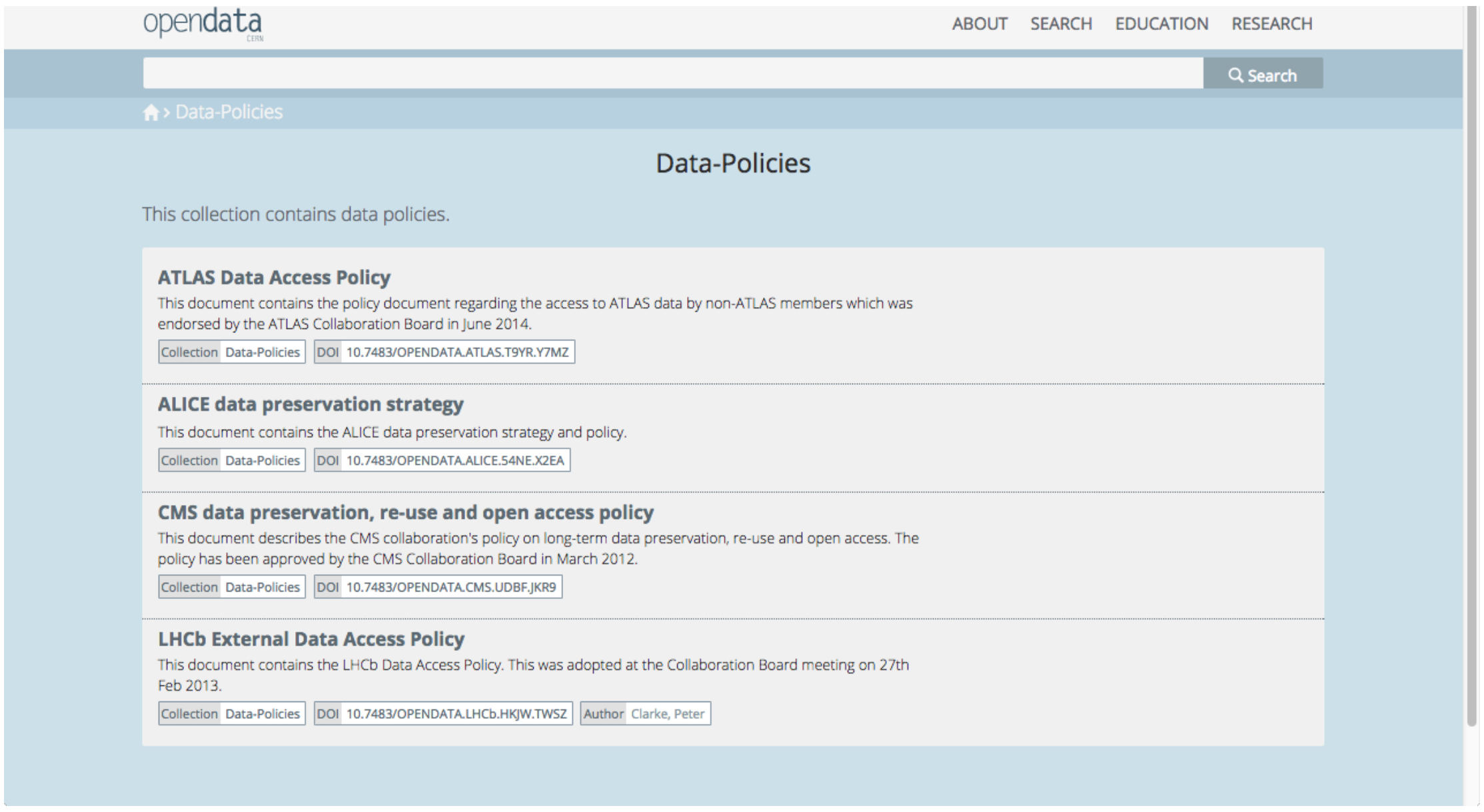

Figure 1 - Data Policies of the Main LHC Experiments

A fraction – up to 100% - of the level 3 data will be shared after an incubation period of several years. Up to date information on the released data can be found via the Open Data portal at http://opendata.cern.ch/.

**Even though this applies to the reconstructed data, the volumes involved could end up being very significant and the technical and financial issues, particularly in the medium to long term (2020+) are not yet understood!**

# DPHEP Portal

First proposed in 2013, the initial idea was to federate the data preservation portals of the various laboratories and institutes involved, providing information on the experiments, data access and release policies, search capabilities and so forth. A much simplified and pragmatic approach is now implemented at: dphep.web.cern.ch
This portal page can be embellished with additional capabilities as manpower allows – in particular for current and future experiments. A simple template is used to provide an overview of the experiment(s) and corresponding accelerator / collider and host laboratory, with drill-down to (largely existing) further detail as needed.

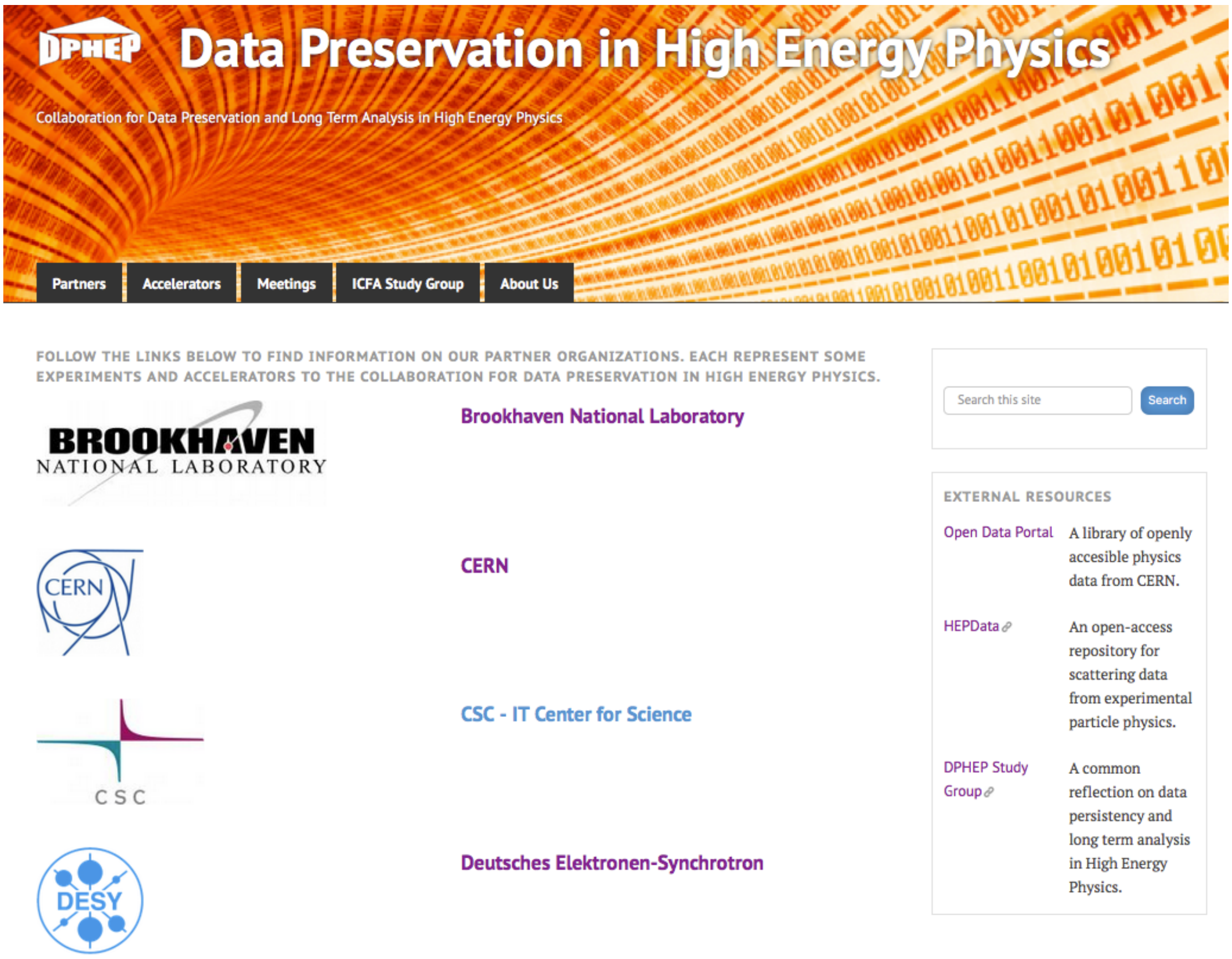

Figure 2 - List of Partner institutions on the DPHEP portal

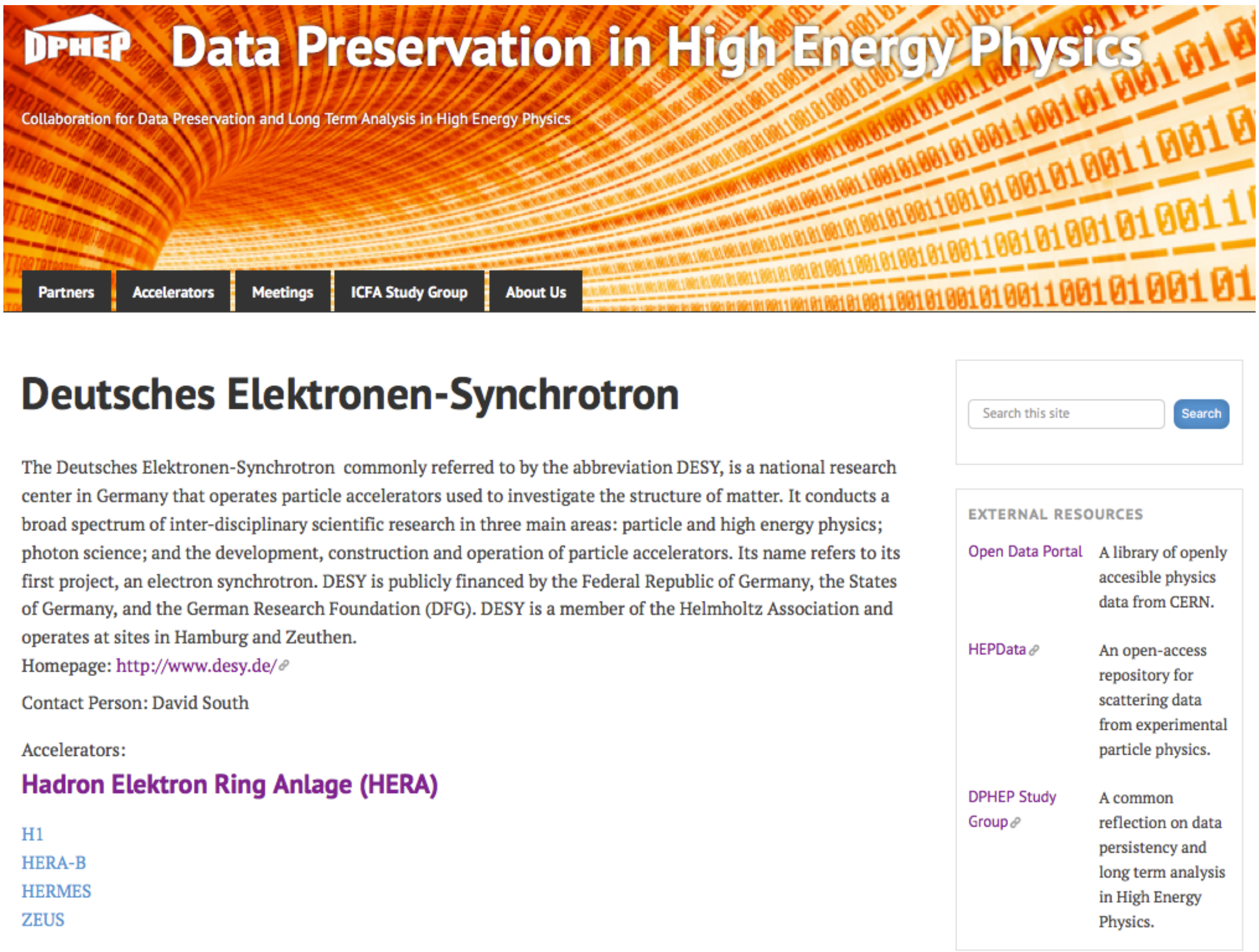

Figure 3 - Example of "Home Page" of an Institute in DPHEP Portal

Information on the data, documentation and software is provided with a standard look and feel, although details are expected to vary.

## The CERN Grey Book

The list of experiments at CERN was published annually from 1975 to 1999 in a printed version of the so-called Grey Book. Since the year 2000 CERN's experimental programme and projects are summarized electronically in the Grey Book database. The information that the Grey Book contains for a given experiment includes:

- A link to the experiments' Website;
- A pointer to the corresponding entry in the CERN Document Server (CDS) in the collection "Experiments at CERN";
- A similar pointer to the CDS collections "Committee Documents" and "Published Articles".

To link the Grey Book to the DPHEP portal (and vice-versa), an additional pointer will be added to point to the Data, Documentation and Software page in the DPHEP portal and a corresponding pointer (e.g. a "grey book" icon) between a given experiment's entry in the DPHEP portal to the Grey Book.

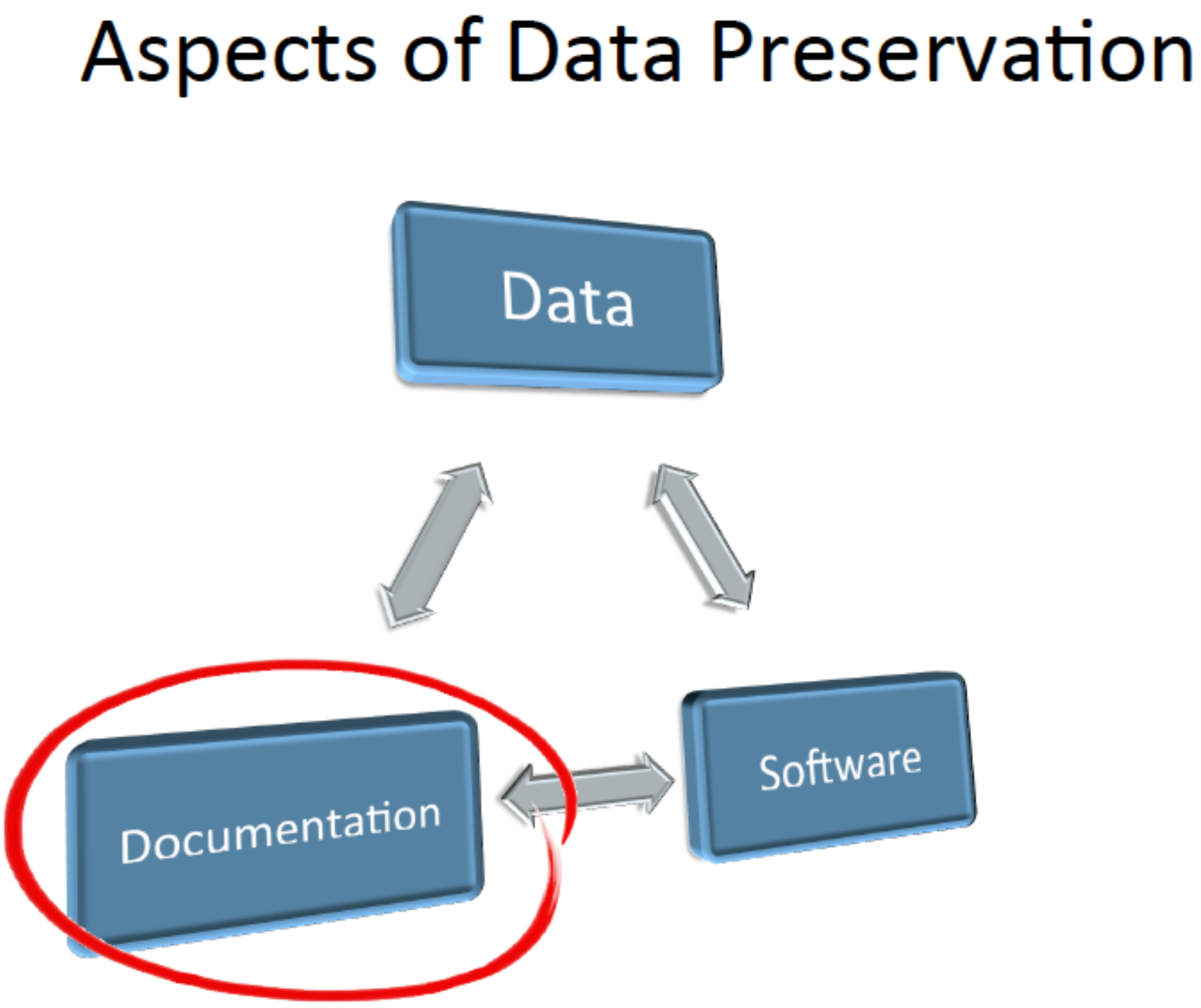

Figure 4 - Key Pillars of Data Preservation

Much of the information should be stable over time, with status reports (e.g. at DPHEP workshops, probably not more than annually) and updates to "HowTos" (updated for e.g. every new operating system release that is supported, changes in data access protocols etc. – hopefully less frequently but probably at least every 3-5 years) being the obvious exceptions.

For programmes such as those at the LHC, links to the analysis capture portal (for those authorised, i.e. collaboration members) and to the open data portal would additionally be provided. Links to external maintained sites – such as the active work on ALEPH data in INFN, that on OPAL data at the Max Planck Institute – would also fit naturally but not disturb the common look and feel.

## Use Cases, Cost Models and Business Cases

Following numerous discussions, a set of common Use Cases has been agreed across the 4 main LHC experiments. With some small provisos, these are also valid for other experiments, including those reported on later in this document.

The basic Use Cases are as follows:

1. Bit preservation as a basic "service" on which higher level components can build;
   - Motivation: Data taken by the experiments should be preserved
2. Preserve data, software, and know-how[14] in the collaborations;
   - Foundation for long-term DP strategy
   - Analysis reproducibility: Data preservation alongside software evolution
3. Share data and associated software with (larger) scientific community
   - Additional requirements:
   - Storage, distributed computing
   - Accessibility issues, intellectual property
   - Formalising and simplifying data format and analysis procedure
   - Documentation
4. Open access to reduced data set to general public
   - Education and outreach
   - Continuous effort to provide meaningful examples and demonstrations

In general, Open Access is not currently considered for pre-LHC experiments that have well defined Open Access Policies. Furthermore, the "designated community" (in OAIS terminology) is typically the (former) collaboration – although there is often considerable flexibility[15] in interpreting this restriction.

These Use Cases map well onto requirements now coming from Funding Agencies for data preservation, sharing and reproducibility. However, it is clear that we will have to work with them to understand and agree on what is technically possible, financially affordable and scientifically meaningful in this area.
A detailed cost model approximating[16] to that for LHC data shows that there is a significant upfront investment that drops rapidly with time. It is based on certain parameters, such as the use of Enterprise tape drives and media for the archive store, together with regular repacking to new, higher density media as this becomes available.

A very simple model that loosely matches the expected evolution in acquired LHC data volumes is shown below.

---

[14] Additional Use Cases – not yet fully tested – help to define whether the "know-how" has been adequately captured. See the Analysis Capture section for further details.
[15] In some cases it is sufficient to join the collaboration (typically by sending an e-mail to the Spokesperson); in others at least one former member of the collaboration must sign any papers and/or an appropriate disclaimer must be included).
[16] The cost model uses publically available pricing information and is thus suitable for sharing with other communities.

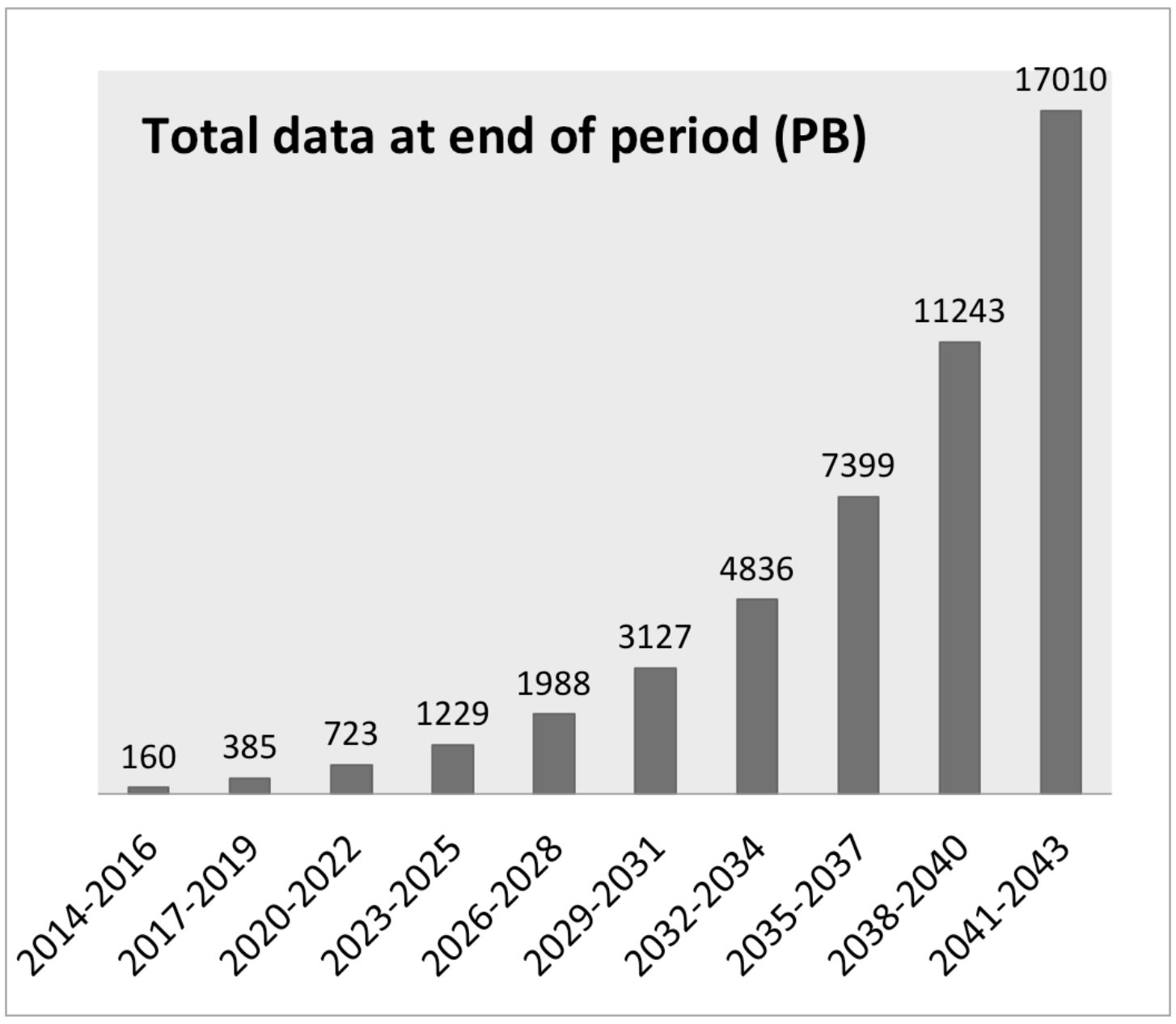

Figure 5 - Approximation to Evolution of LHC Storage for Cost Model

Based on publically available technology predictions and pricing information, we are able to calculate how much it would cost to store a single copy this information in a set of tape libraries (a 10% disk cache is included, as is a 3-year cycle for the media, after which all data are migrated forward to the next generation).

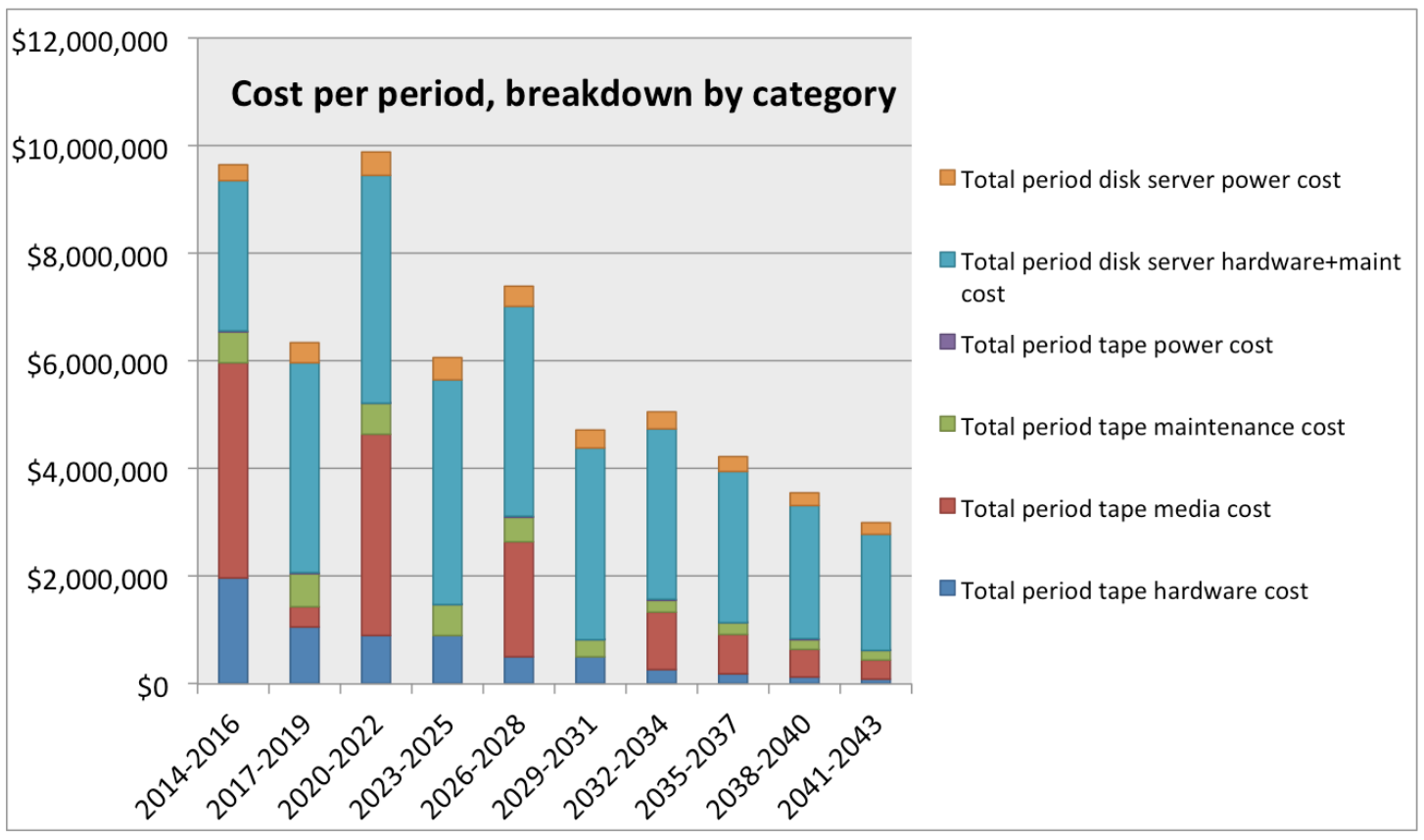

Figure 6 - Breakdown of Costs According to Storage Growth / Media Replacement

Not only does this – together with on-going data scrubbing – implement some of the key practice in OAIS and the associated certification procedures and hence allow us to offer "state-of-the-art" bit preservation, but we can also calculate the costs of a data store rising from several tens of PB initially to a few EB in the 2030s. Whilst much more detailed calculations are used in the LHC (WLCG) budget review and request process, this gives us at least a ballpark estimate for the costs involved and we can see that the cost over time averages to "just" $2M / year (for such a vast and growing data store).

Comparatively, e.g. versus the cost of LHC computing, the cost of building and running the machine and its detectors, this is a "small number" – certainly much less than the cost of building a new machine in the future (at least with today's technology)!

The "value" of the preserved data can be measured indirectly by the number of on-going analyses, publications and / or major conference presentations, as shown in the figures below for CDF, D0 and *BABAR*. These all show that there is significant activity that continues well after the end of data taking.

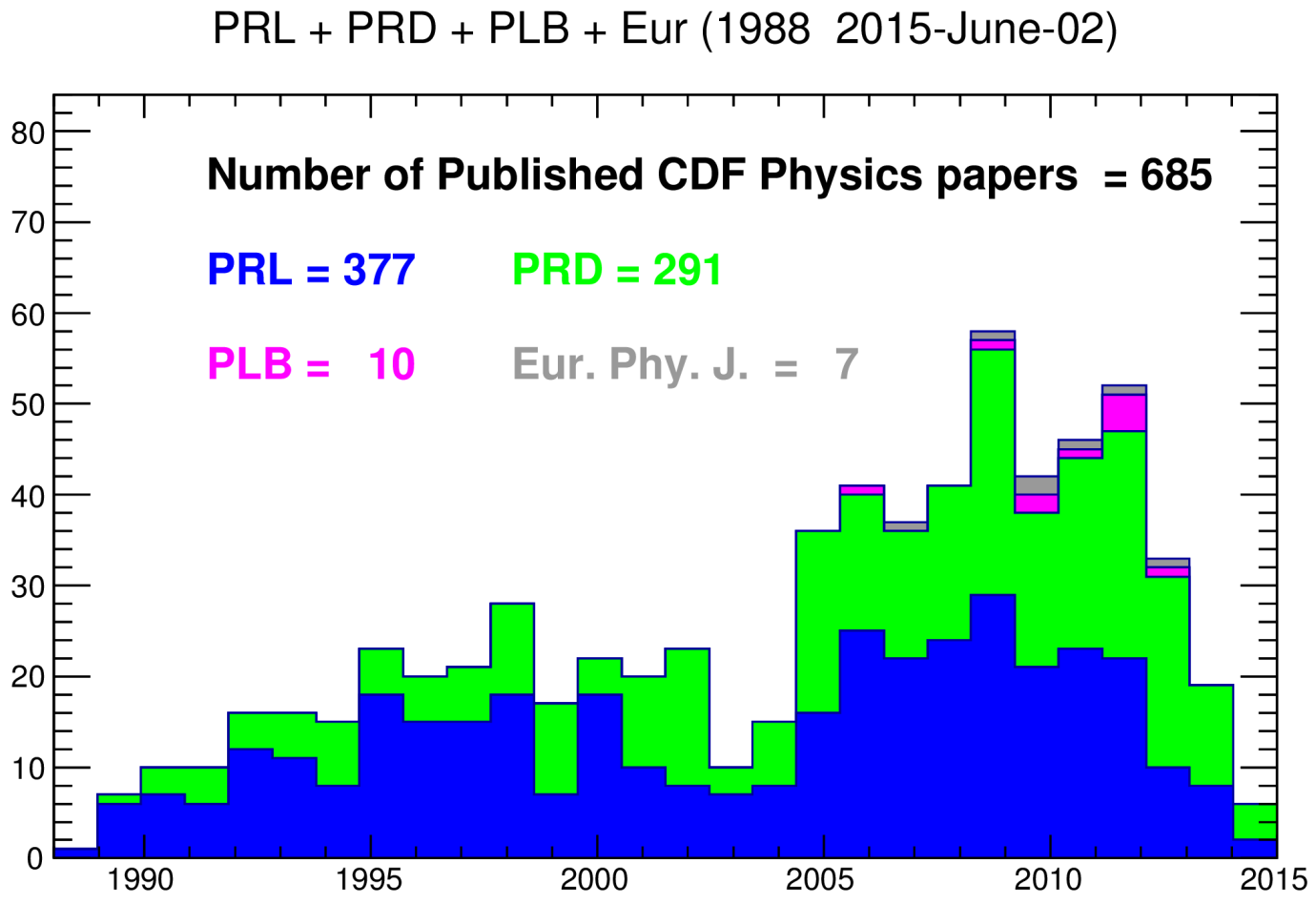

Figure 7 - Published Papers for the CDF Collaboration

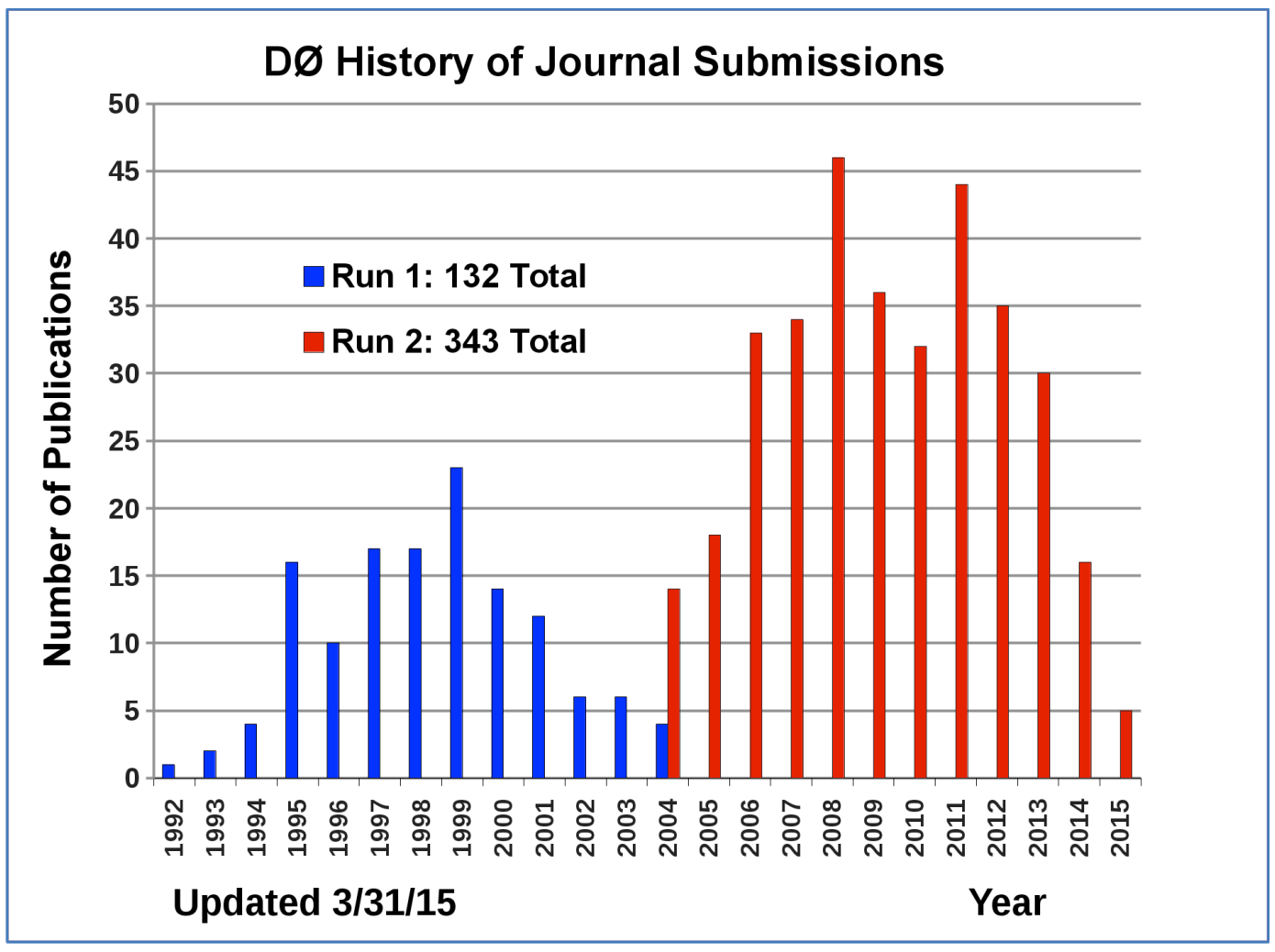

Figure 8 - Journal Submissions for the D0 Experiment

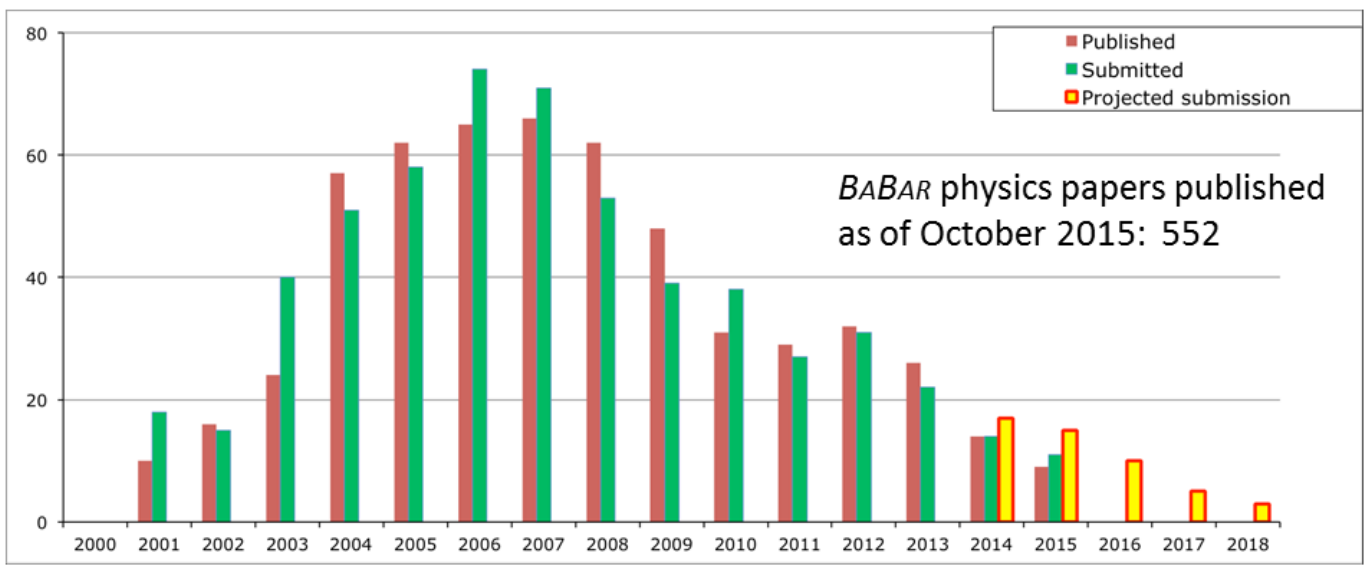

Figure 9 - Papers Submitted / Published for *BABAR*

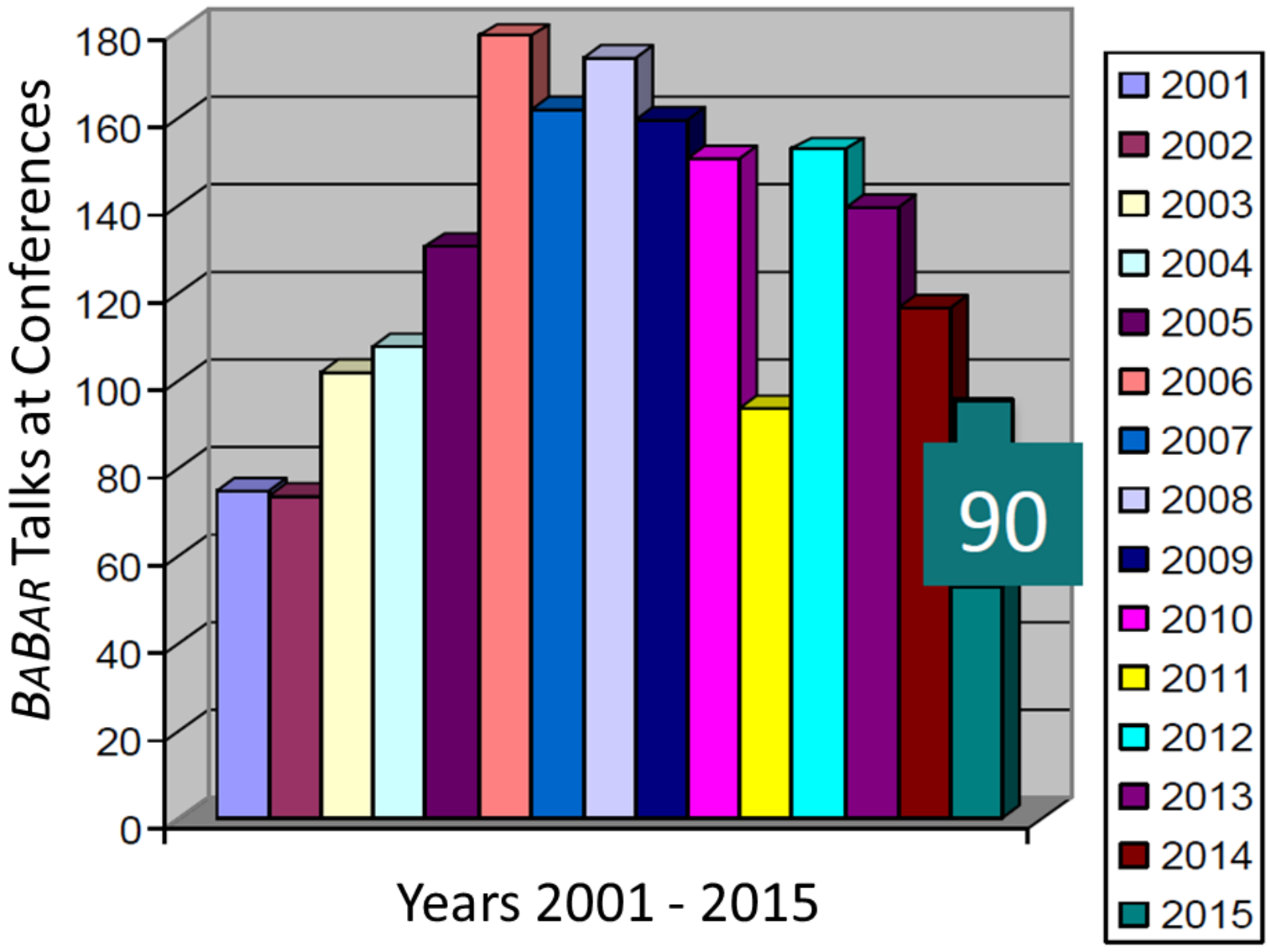

Figure 10 - Conference Talks by Year for *BABAR*

# Bit Preservation and Storage Technology Outlook

Bit preservation is an art in itself and – following the 4C project recommendations (see below) – is best performed at a limited number of “expert” sites, rather than across a multitude of smaller ones. This becomes even more important as densities increase – whereas user manipulation of individual tape volumes was common place in the LEP era, the latest generations of media requiring extreme clean-room conditions and prefer robots over humans!

The following graph shows the growth in data stored at CERN for the LHC and other experiments.

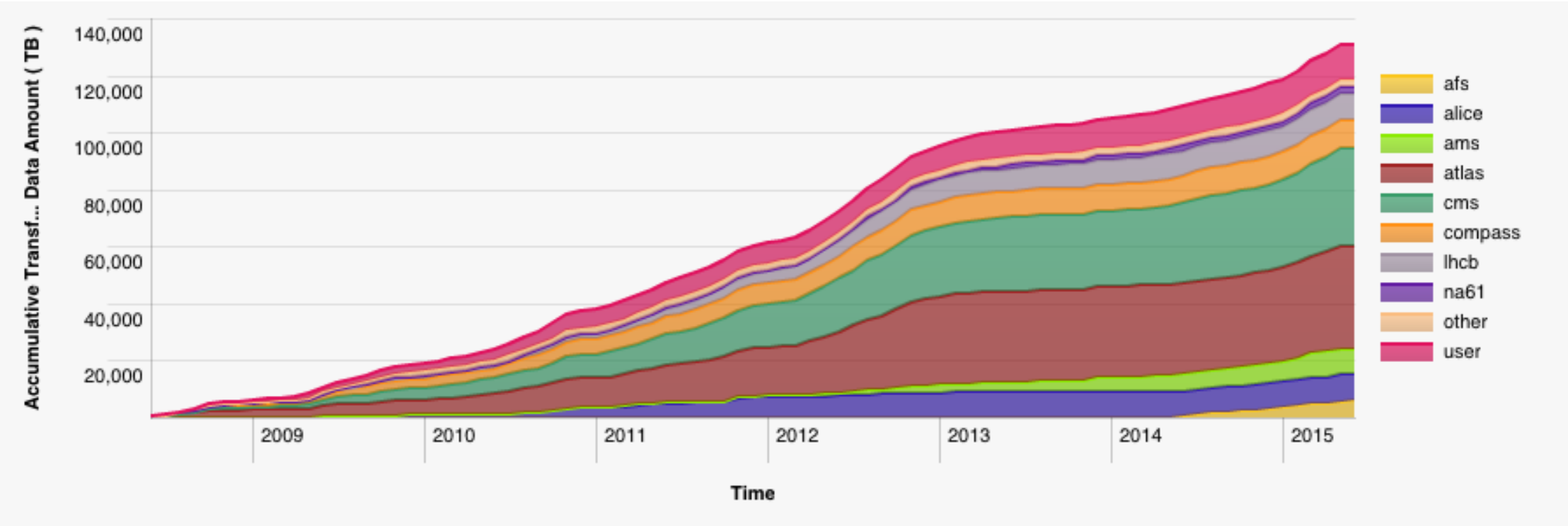

Figure 11 - Growth of Experiment (and AFS) data stored at CERN

Based on the anticipated data rates and volumes at LHC Run2 and future running periods, we predict a total data volume of a few EB (exabytes) in the 2030s.

Industry predictions (see below) suggest that cartridge capacity can continue to grow, at least over the next few years.

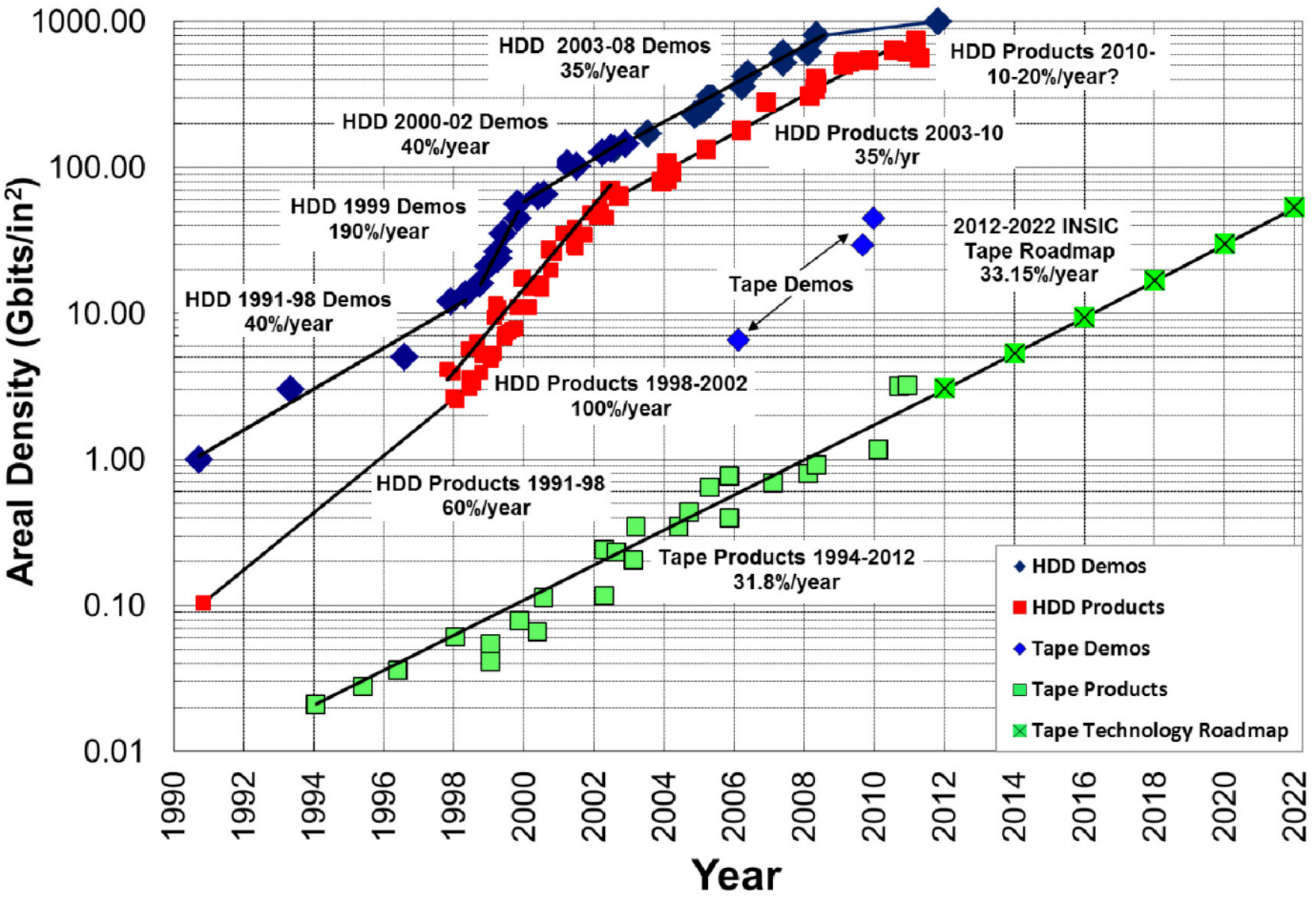

Figure 12 - Expected Evolution in Disk and Tape Technology

However, one has to inject a word of caution here – the tape market is shrinking, the source of enterprise drives has become a duopoly, with but a single supplier of high-density media.

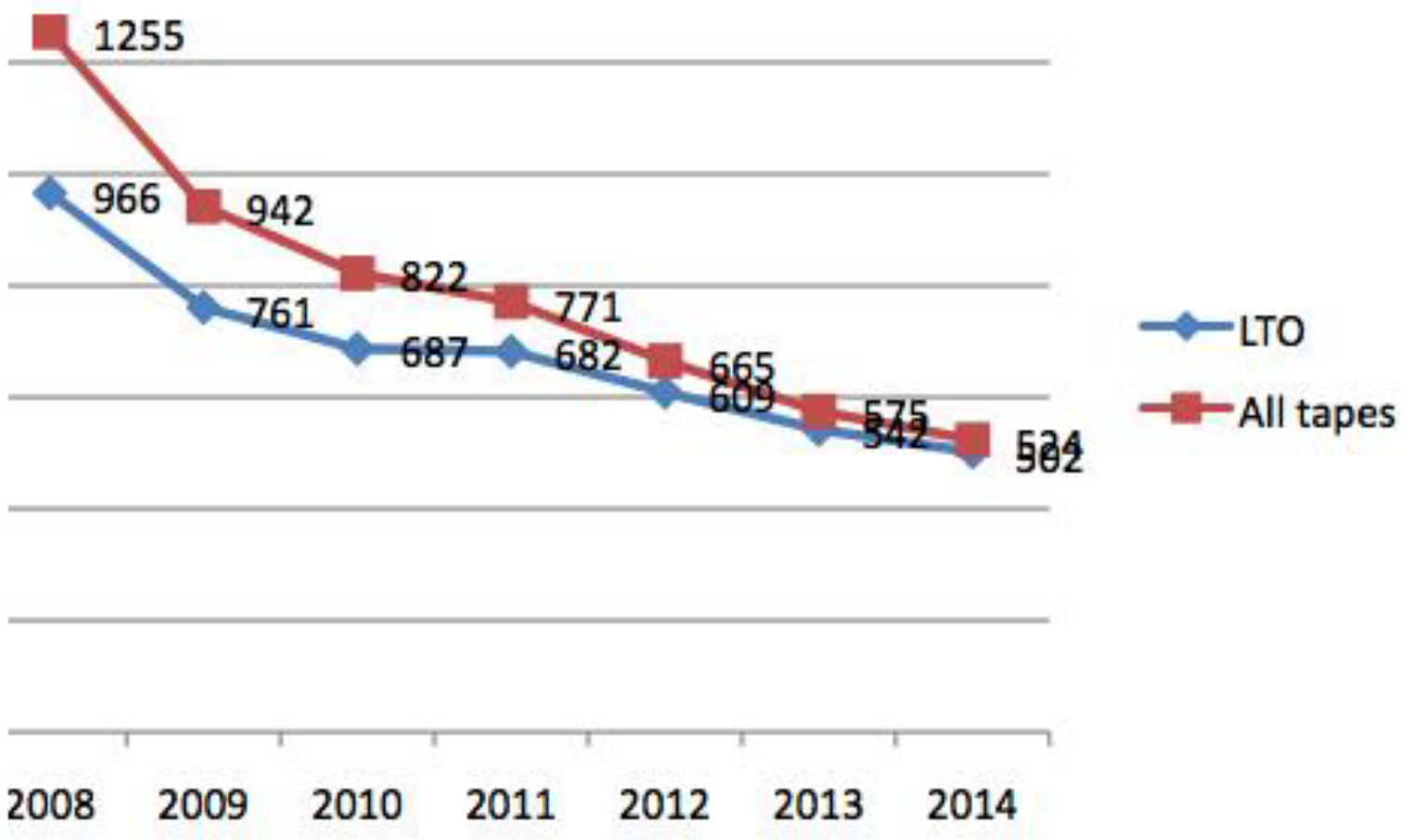

Figure 13 - Evolution of Tape Market

Some experts – such as David Rosenthal[17] - predict that Kryder's law (the "equivalent" of Moore's law for storage) will no longer hold true in the future. He warns that we should expect to pay more for storage. There have been many predictions of storage revolutions in the past – often involving optical or holographic storage. However, these have so far failed to materialise.

Looking back, we see a significant improvement in storage capacity with the number of cartridges required to storage all LEP data shrinking to an almost negligible number. So much so that now two tape copies are maintained at CERN, with a further read-only disk copy being setup. Given the additional copies maintained at a number of outside institutes for at least some of the LEP experiments, we have achieved a significant level of redundancy. Will this be true one day also for the LHC experiments?

[17] See http://blog.dshr.org/.

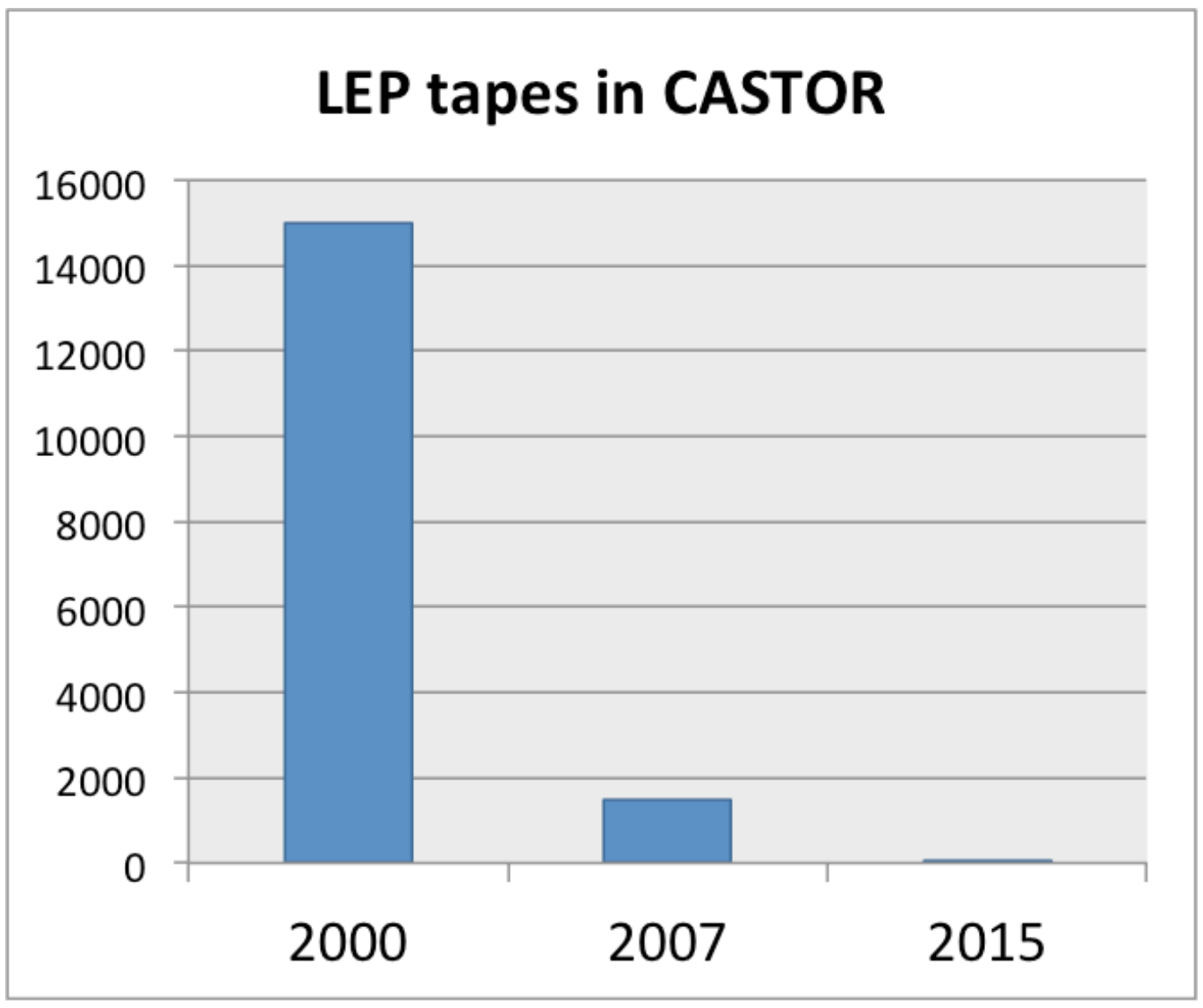

Figure 14 - Number of Tapes Required to Store LEP data

We can also see a measurable improvement in terms of reliability at the level of a single site. Additional levels of protection are foreseen, hopefully reducing data loss further still.

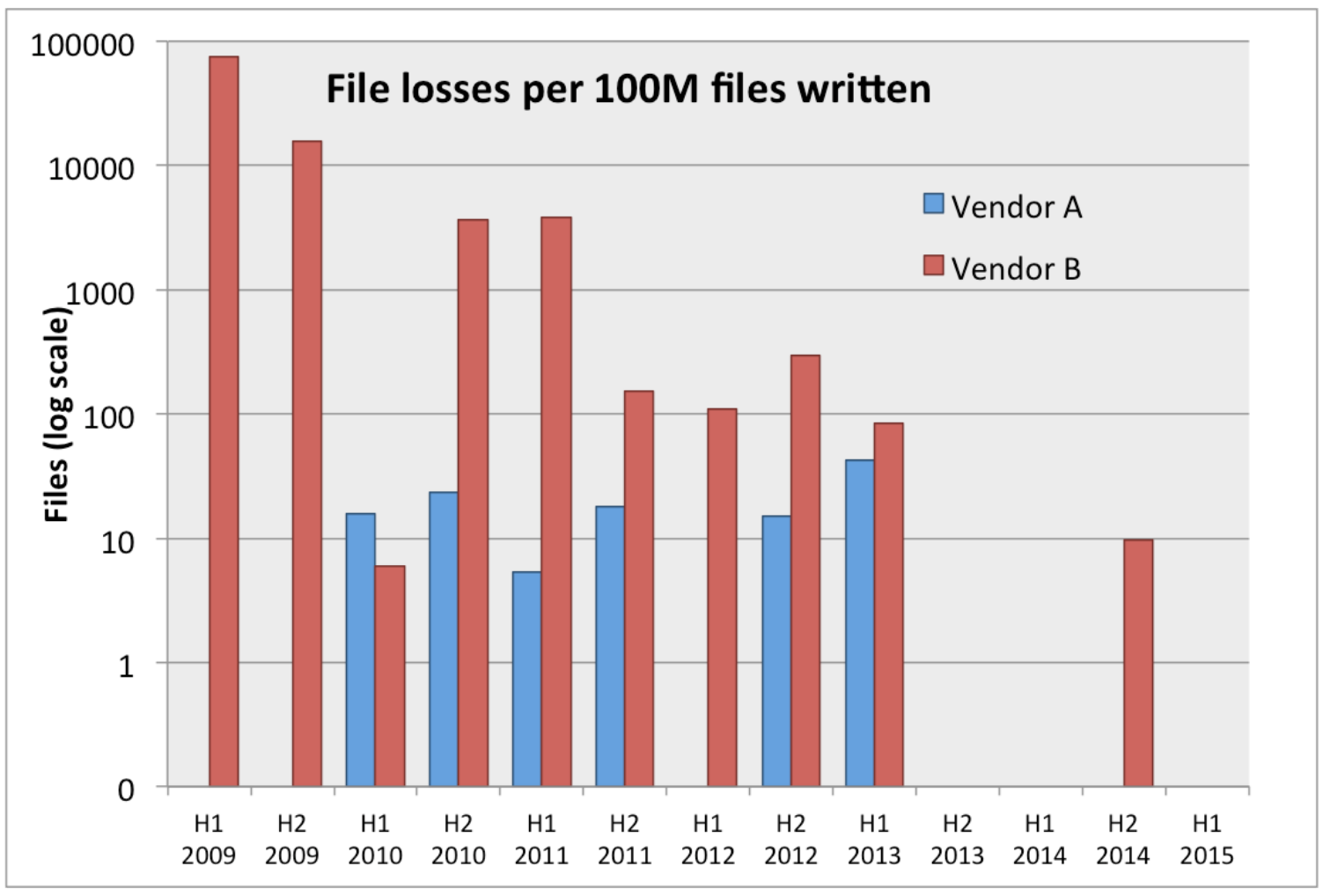

Figure 15 - File Losses per 100M files written

This work can be compared to the recommendations of the National (i.e. US) Digital Stewardship Alliance (NDSA), shown in the table below.

| | Level 1 (Protect your data) | Level 2 (Know your data) | Level 3 (Monitor your data) | Level 4 (Repair your data) |
|---|---|---|---|---|
| **Storage and Geographic Location** | • Two complete copies that are not collocated<br>• For data on heterogeneous media (optical discs, hard drives, etc.) get the content off the | • At least three complete copies<br>• At least one copy in a different geographic location<br>• Document your | • At least one copy in a geographic location with a different disaster threat<br>• Obsolescence monitoring process for your storage system(s) and | • At least three copies in geographic locations with different disaster threats<br>• Have a comprehensive plan in place that |
| | medium and into your storage system | storage system(s) and storage media and what you need to use them | media | will keep files and metadata on currently accessible media or systems |
| **File Fixity and Data Integrity** | • Check file fixity on ingest if it has been provided with the content<br>• Create fixity info if it wasn't provided with the content | • Check fixity on all ingests<br>• Use write-blockers when working with original media<br>• Virus-check high risk content | • Check fixity of content at fixed intervals<br>• Maintain logs of fixity info; supply audit on demand<br>• Ability to detect corrupt data<br>• Virus-check all content | • Check fixity of all content in response to specific events or activities<br>• Ability to replace/repair corrupted data<br>• Ensure no one person has write access to all copies |
| **Information Security** | • Identify who has read, write, move and delete authorization to individual files<br>• Restrict who has those authorizations to individual files | • Document access restrictions for content | • Maintain logs of who performed what actions on files, including deletions and preservation actions | • Perform audit of logs |
| **Metadata** | • Inventory of content and its storage location<br>• Ensure backup and non-collocation of inventory | • Store administrative metadata<br>• Store transformative metadata and log events | • Store standard technical and descriptive metadata | • Store standard preservation metadata |
| **File Formats** | • When you can give input into the creation of digital files encourage use of a limited set of known open formats and codecs | • Inventory of file formats in use | • Monitor file format obsolescence issues | • Perform format migrations, emulation and similar activities as needed |

Figure 16 - NDSA Levels of Digital Preservation

# Virtualisation and Software Preservation

In order to process the data of complex scientific instruments such as particle detectors, scientists use and develop complex software systems. These software systems, for instance, simulate the response of the instruments to physics processes under study, they reconstruct true physics information from raw detector signals, they interpret and convert data formats, they can visualize data sets and they provide mathematical routines and machine learning frameworks for the statistical analysis of the data. Having large international collaborations, high-energy physics has a long tradition of sharing and developing common, open-source software stacks. These stacks are composed from industry standard software (such as Linux, compilers, mathematical libraries), high-energy physics specific software (such as the ROOT and Geant4 toolkits) and software specific to an experiment or a research topic. For the LHC experiments, the software stacks accumulate tens of millions of lines of code, half a dozen different languages and tens to hundreds of modules with dependencies on each other that need to be configured to act as a coherent system. Furthermore, the petabytes of data recorded by modern instruments require data processing tasks to scale out to distributed systems for the mid-term future (5-10 year at least). That involves an additional software stack of middleware used for data bookkeeping, data distribution, resource control, and user authorization.

The porting and validation of such complex software stacks to contemporary technologies is in many cases an effort that shrinking collaborations of decommissioned experiments cannot afford. Hardware virtualization (such as VMware and VirtualBox) and container virtualization (such as Docker) provide a resort as they allow a frozen, historic software environment being executed on contemporary hardware and operating systems.

In a naive application of virtualization technology, a software environment is frozen in the form of a disk image, a large and opaque stream of bytes containing all the necessary software binaries. This approach tends to be clumsy and too rigid for HEP software. In order to be useful, even "frozen" environments need to stay open for minor modifications: patches to faulty algorithms, updates to physics models, updated tuning parameters for simulation algorithms, new configuration for data access software and so on. Software development communities have long solved similar problems by version control systems. Version control systems only store a track of changes to the source code and at the same time they can provide access to the state of a directory tree at any given point in the past.

CernVM and the CernVM File System are open source technologies developed and maintained at CERN that provide a portable software development and runtime environment for HEP experiments. They are based on virtualization technology and versioning file system technology such that virtualization itself is separated from the concerns of accessing software binaries. A minimal and stable virtual machine or container (<20MB) connects to a remote file system (CernVM-FS) that maintains a versioned repository of the operating system and software binaries. By selecting different states of the versioned file system, we use CernVM to create software environments compatible with Red Hat Enterprise Linux (RHEL) 4 to RHEL 7 (spanning 10+ years) with the very same virtual machine on the very same hardware.

Both technologies are used in production by LHC experiments and other scientific collaborations. The scale and criticality for active experiment collaborations ensure support and maintenance in the long run. CernVM-FS is supported on all major Linux flavours and OS X. It is deployed on 65,000+ physical machines distributed worldwide and it hosts several hundred million files and directories of experiment software. CernVM runs on all major cloud computing platforms, including commercial platforms Amazon EC2, Google Compute Engine, and Microsoft Azure. There are 3000+ interested citizens (volunteers) that run CernVM on their computers in the context of LHC@Home 2.0 and there are more than 1 million CernVM virtual machines (re-)booting every month.

Three different use cases demonstrate the applicability of CernVM and CernVM-FS for data preservation.

1) The CMS Open Data Pilot: In this exercise, CernVM creates a CMS software environment that dates back to 2010. This particular software version is validated for the data set from 2010 that is provided to the general public. There was no additional effort required to restore the historic CMS software as CMS software is automatically stored and versioned in CernVM-FS for everyday use. The CMS Open Data Pilot is part of the CERN OpenData Portal.

2) The ALICE Open Data VM: In this exercise, contemporary ALICE software is used to provide access to the publicly released data set of ALICE. It demonstrates that, properly packaged, the current live software of an experiment can be made available to interested citizens for outreach and education with very little cost.

3) ALEPH software in CernVM: In this exercise, ALEPH software was installed post-ex on CernVM-FS and made available in a RHEL 4 compatible CernVM on the current CERN OpenStack infrastructure. It demonstrates that the technology is able bridge technology evolution over 10+ years.

In order to better separate the needs for preserving scientific applications from the generic data access middleware, current developments aim at using both hardware virtualization and container virtualization. Contemporary virtual machines provide data access tools and middleware with support for the latest network protocols and security settings. Containers inside the virtual machines spawn historic operating system and application software environments. Data is provided from the container host to the historic applications through the very stable POSIX file system interface.

# Documentation and Digital Library Technologies

The technical know-how necessary to operate the complex software stack and understand the data content of each HEP experiment is captured in documents. These documents are:

- Technical manuals usually written using LaTeX and available in portable document format (PDF), postscript (PS), hypertext markup language (HTML).
- Operational information often captured in wikis and electronic logbooks. These usually use TWiki and elog as their technical platforms.
- Information exchanged during meetings, which is contained in MS PowerPoint or PDF slides and text minutes, as well as video recordings.

# CERN Program Library Documentation and Software

Many HEP experiments rely to a greater or lesser degree on the set of libraries known collectively as the "CERN Program Library" or simply CERNLIB. The documentation for these was last revised in the mid-1990s with the sources, marked up in LaTeX, stored at CERN in /afs.

In order to best preserve the documentation for the medium to long term, the following activities have been performed:

1. Reformatting of the source files to produce PDF and/or PDF/A files with the latest fonts;

2. Capturing of the author and paper information, storing of the formatted files in the CERN Document Server using identifiers to refer to the authors and papers;
3. Addition of further meta-data to enable more powerful searches;
4. Storing of the resultant files in the CERN Document Server (CDS) – see below.

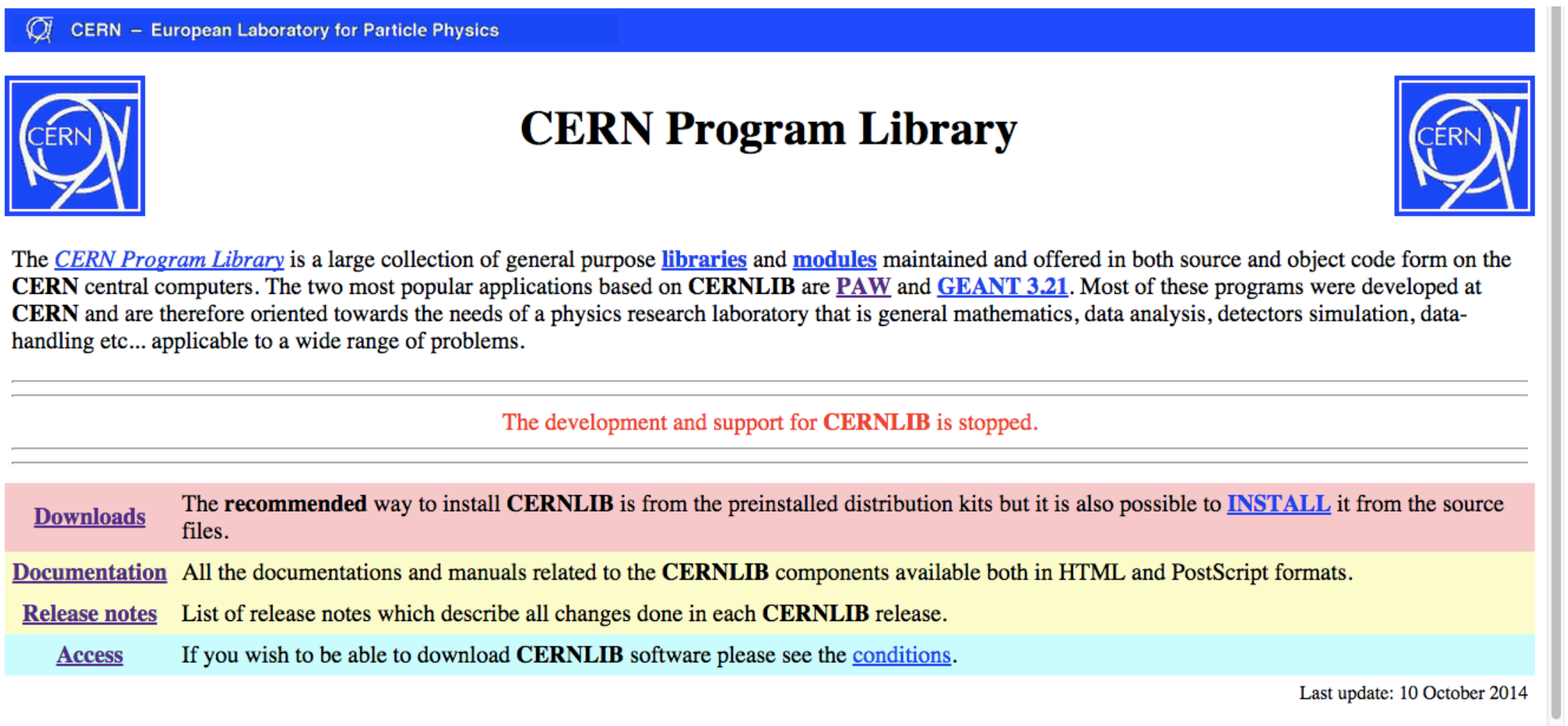

Figure 17 - "home page" for CERNLIB

Formal support for CERNLIB ceased over a decade ago – and development earlier still. However, it continues to be actively used in “data preservation” and re-use activities. Porting to future versions of Linux and an “official” version that the (past) experiments can trust is still desirable.

The “long” (typically complete packages such as HBOOK, PAW or ZEBRA) and “short” (mainly individual or groups of routines from KERNLIB and MATHLIB) write-ups of the CERN Program Library have been recompiled from their LaTeX sources. PDF and HTML versions of the documents were created. The short write-up documents have been added to the CERN Document Server (CDS). The CDS records are described by metadata keywords derived from the descriptions embedded in the LaTeX source and the write-up text. The collection of records is available under the “Software Documentation” section of CDS: an example is shown below.

| CLOGAM | CERN Program Library | C333 |
|---|---|---|

**Author(s) :** K.S. Kölbig — **Library:** MATHLIB
**Submitter:** — **Submitted:** 10.04.1972
**Language :** Fortran — **Revised:** 15.03.1993

**Logarithm of the Gamma Function for Complex Argument**

Function subprograms CLOGAM and WLOGAM calculate the logarithm of the gamma function

$$\ln\Gamma(z)$$

for complex $z \neq -n$, $(n = 0, 1, 2, \ldots)$. The imaginary part $\operatorname{Im} \ln\Gamma(z)$ is calculated in such a way that it is continuous for $|\arg z| < \pi$, i.e. it is not taken mod $(2\pi)$.

The double-precision version WLOGAM is available only on computers which support a COMPLEX*16 Fortran data type.

**Structure:**

FUNCTION subprograms
User Entry Names: CLOGAM, WLOGAM
Files Referenced: Unit 6
External References: MTLMTR (()N002), ABEND (()Z035)

**Usage:**

In any arithmetic expression,

CLOGAM(Z) or WLOGAM(Z) has the value $\ln\Gamma(\mathrm{Z})$,

where CLOGAM is of type COMPLEX, WLOGAM is of type COMPLEX*16, and Z has the same type as the function name.

**Method:**

The method is described in Ref. 1.

Figure 18 - Extract of a reformatted CERNLIB Short Writeup

# Analysis Preservation

Repositories make a good match for long-term preservation needs. However, during their active lifetime, experiments make use of many other systems and produce a wide range of objects relevant to the research outcome. These dynamic products need to be met by new data preservation and Open Science services. In the past it had been complex, if not impossible, to preserve (and share) such research objects. This resulted in challenges when compiling information for analysis or publication approval or when revisiting an analysis after it had been finished for a long time. This is by no means a HEP-specific problem; it applies equally to all disciplines. Hence, the aim is to capture all digital assets and associated knowledge inherent in the data analysis process for subsequent generations and to make a subset available rapidly to the public. Additionally, tools for keeping a modern, electronic, logbook, are of prime importance. Thus CERN Analysis Preservation (CAP) was launched as a tool for data preservation while the analysis is still active, enabling researchers to preserve the selected content and related information needed to understand the analysis.

CAP is now ready as a first prototype. It helps the data preservation process by capturing information from the very beginning of the research idea, e.g. through a connection with the respective job databases in a particular collaboration. This will allow the capturing of all the changes made to the code used for analysis and thus be similar to an electronic logbook. An open, annotated and structured electronic logbook would allow data and its documentation to be found easily. Moreover, this enables collaborations or specific groups to integrate CAP into approval workflows, e.g. for analysis, conference or publications. Also, CAP will provide search options that could allow newcomers to get an overview of the analyses that are already on-going or could be finished more easily. Moreover, CERN Analysis Preservation will help to resolve research conflicts by providing detailed information on the analysis.

The pilot solution of CAP has been prototyped using the Invenio digital library platform, which was extended with several data-handling capabilities. The aim was to preserve information about datasets, the underlying OS platform and the user software used to study them. The configuration parameters, the high-level physics information such as physics object selection, and any necessary documentation and discussions are optionally being recorded alongside the process as well. The metadata representation of captured assets uses the MARC bibliographic standard, which had to be customised and extended in relation to specific analysis-related fields. The captured digital assets are being minted with Digital Object Identifiers (DOIs), ensuring later referenceability and citability of preserved data, documentation and software.

CAP will be offering APIs to enable easy information exchange between CAP, CODP, INSPIREHEP, CDS and other, possibly external, platforms. Thus, upon user request, data can be published openly from CAP through the partnering open portals.

The ultimate goal of the CERN Analysis Preservation platform is to capture sufficient information about the process in order to facilitate reproduction of an analysis even many years after its initial publication, permitting extension of the impact of preserved analyses through future revalidation and recasting services.

## CERN Open Data

The CERN Open Data portal (CODP) was launched in November 2014. The CODP service is the access point to a growing range of data produced through the research performed at CERN. Currently the portal publishes public data releases from the CMS, ALICE, ATLAS and LHCb experiments. The data are accompanied by software and documentation that is needed to understand and analyse the data being shared.

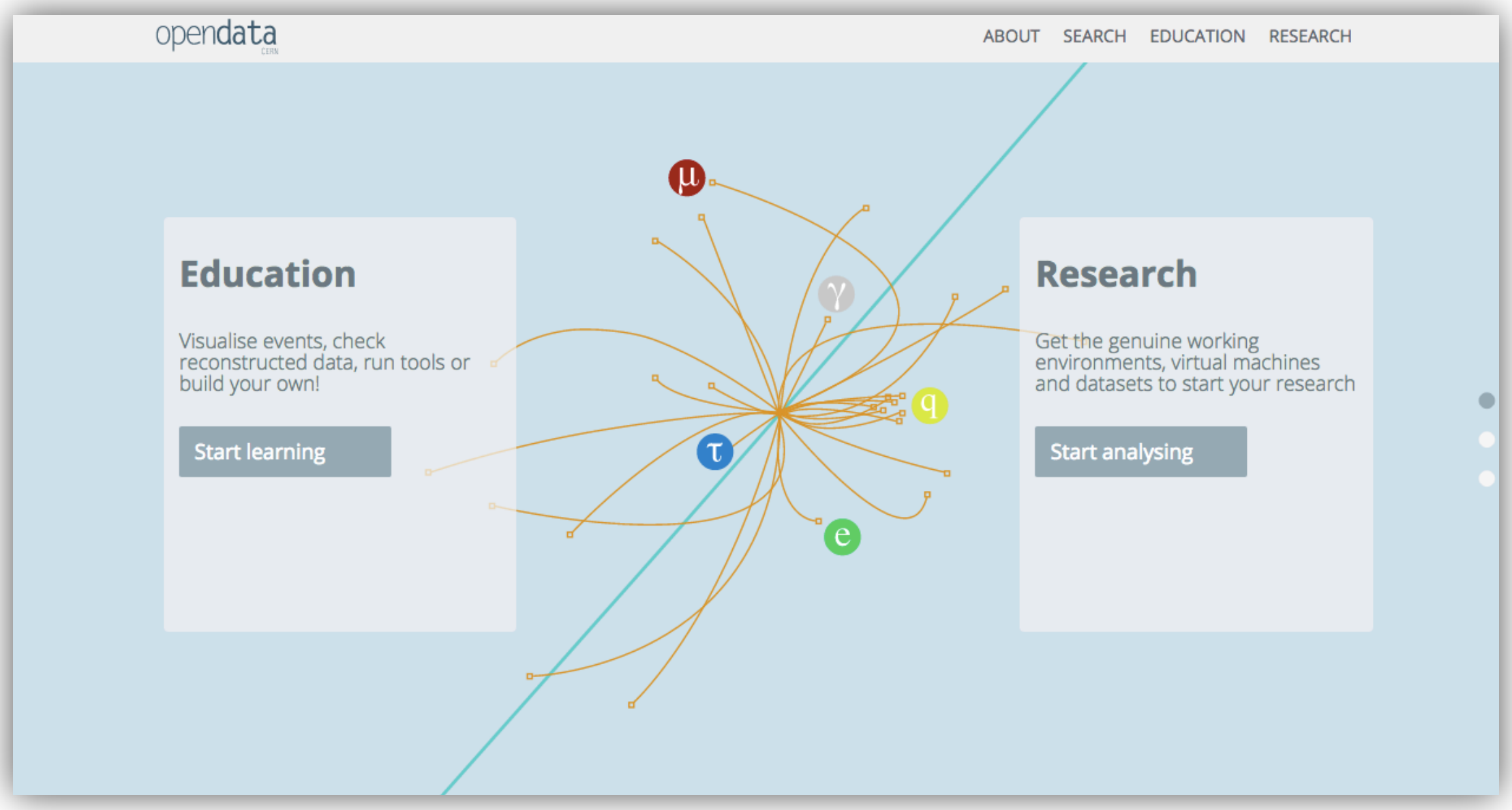

Figure 19 - the CERN "Open Data" Portal

The CODP offers two entry points, namely *Education* and *Research*. Categorizing outputs in this way should make it easier for the target groups to identify materials relevant to their interests.

In *Education*, users can access simplified data formats for analysis in outreach and training exercises. These simplified data formats are provided with a comprehensive set of supporting material, so that they could easily be used by high-school teachers and students in CERN's masterclasses, for example. Students can use datasets, reconstructed data, tools and learning resources to further explore the data and improve their knowledge of particle physics. Therefore, a considerable part of the CERN Open Data portal was devoted to tailored presentation and ease-of-use of captured data and associated information.

In *Research*, the portal presents datasets for research, including reconstructed data, needed software, an example analysis, and guides for virtual machines. The datasets being shared here are explained in detail, e.g. how they have been validated and how they could be potentially used. CODP offers the download of Virtual Machine images, permitting users to start their own working environment in order to further explore the data. For this the platform uses CernVM-based images prepared by the collaborations. With the initial release, CODP provides access to the big data release by the CMS collaboration as noted in the collaborations data policy, also contained in the *Research* part. All the other LHC collaborations joined the release with shared data for education purposes. All LHC collaborations have approved data preservation, re-use and open access policies, which are available on the portal. According to CODP data policies the collaborations are committed to preserving their data, to allow their re-use and to make their data open after a certain embargo period to the wider scientific community and the public.

In conclusion, CODP (with its integration with CAP) provides a structure for data management plans and a focal point for preservation actions. The portals adhere to established global standards in data preservation and Open Science. CODP uses the MARC bibliographic standard to describe the records with metadata for the purpose of discovery and identification. In addition, citability, archiving and preservation are supported, as metadata ensures that data and associated information will survive and continue to be accessible into the future. Consequently, the products are shared under open licenses (Creative Commons CC0 waiver) and they are issued with a DOI to make them citable objects in the scientific discourse. The CERN Open Data Portal endorses the FORCE 11 Joint Declaration of Data Citation Principles, so the data provided in the portal can be cited when they are re-used.

## INSPIREHEP

INSPIREHEP is the main information management system and the main literature database for HEP. It has been produced by a collaboration of five key labs in the HEP field: CERN[18], DESY[19], Fermilab[20], IHEP[21] and SLAC[22].

Originally based on the traditional role as an aggregator for scholarly content, INSPIREHEP has since moved into research data as well. Through a strong partnership with HEPData, data underlying publications are now accessible via the respective publications.

INSPIRE will expand upon this partnership by launching a data collection covering the main data platforms in HEP, from HEPData to CODP. The INSPIRE data collection aims at making public datasets from the HEP community searchable and the content more easily discoverable.

It will be ensured that all datasets are assigned a persistent identifier and data citation will be encouraged. Already now INSPIRE is able to track data citation. In the future, it will be investigated how data citation can be included in author profile pages and how INSPIRE presents data citation counts.

By offering these new services, INSPIREHEP, together with CAP and CODP closes a loop in data preservation, by adding a discovery layer to the existing suite of services. Data preservation is supported from the beginning of the research workflow to the final publications, including referencing and potential reuse.

## HEP Software Foundation

The HEP Software Foundation (HSF)[23] facilitates coordination and common efforts in High Energy Physics (HEP) software and computing internationally. The objectives of the HSF as a community-wide organization include

- Sharing expertise;

[18] See http://home.cern/
[19] See http://www.desy.de/
[20] See http://www.fnal.gov/
[21] See http://english.ihep.cas.cn/
[22] See https://www6.slac.stanford.edu/
[23] See http://hepsoftwarefoundation.org/.

- Raising awareness of existing software and solutions;
- Catalysing new common projects;
- Promoting commonality and collaboration in new developments to make the most of limited resources;
- Aiding developers and users in creating, discovering, using and sustaining common software;
- Supporting career development for software and computing specialists;
- Provide a framework for setting goals and priorities, and attracting effort and support;
- Facilitate wider connections with other science fields.

Although not directly related to data preservation, it is clearly a forum where long-term sustainability of software can be discussed and contacts have been established with this in mind.

## Related Projects, Disciplines and Initiatives

The European Alliance for Permanent Access (APA) was set up as a non-profit organization, initiated as a Foundation under Dutch Law in September 2008. The goal of the Alliance is to align and enhance permanent information infrastructures in Europe across all disciplines. It is a networking organisation and a sustainable centre for advice and expertise on permanent access. The Alliance brings together seventeen major European research laboratories, research funders, and research support organisations such as national libraries and publishers. All its members are stakeholders in the European infrastructure for long-term preservation of and access to the digital records of science. Through the alliance, they are articulating a shared vision for a sustainable digital information infrastructure providing permanent access to scientific information.

CERN has been a member of the APA for many years, as well as having a seat on the Executive Board. Unfortunately, due to an unfavourable audit of an FP7 project in which the APA was involved, it is in the process is dissolution.

The APA held regular conferences where CERN and other partners presented. It also played a key role in FP7 projects including APARSEN, SCIDIP-ES, PRELIDA and others.

Through contacts with the APA, the activities of DPHEP have become much more widely known outside HEP, as well as specific activities such as peta- to exa-scale bit preservation and cost modelling.

A series of joint workshops have been held between the above projects and DPHEP at several meetings of the Research Data Alliance (RDA).

Material from these joint workshops can be found through the DPHEP Indico pages: https://indico.cern.ch/category/4458/.

The RDA (https://rd-alliance.org/node) is now three years old, is supported by funding agencies in North America, Europe and Asia-Pacific and has a focus on data sharing and re-use. It holds two plenary meetings per year that include working meetings of numerous Working and Interest groups. (Working groups are supposed to deliver tangible outputs in around 18 months whereas interest groups are longer lived and are one mechanism by which working groups can be setup). Groups of particular interest to DPHEP include:

- The Preservation e-Infrastructure Interest Group;
- WG and IGs on (harmonization of) certification of digital repositories;
- Active Data Management;
- Reproducibility;
- Citation;
- And so forth.

As a networking event the RDA meetings can be particularly valuable and contacts made through the RDA as well as APA have helped establish and refine our vision, as well as providing channels whereby our activities can be more widely disseminated.

Furthermore, the RDA appears to be a central point for discussing "all things data" and clearly has the attention of the funding agencies. The RDA Europe arm of the "project" will likely be funded (in several stages) throughout the entire Horizon 2020 programmes. CERN – on behalf of the EIROforum IT Working Group – is currently an Organizational Member and has a seat on the Technical Advisory Board (September 2013 – September 2015). It is also represented on the advisory board of the RDA Europe H2020 project.

### EU FP7 Projects

Three FP7 policy projects – 4C, RECODE and PERICLES – are also worthy of discussion, as summarized below.

## 4C and RECODE Policy Recommendations

**4C** was an FP7 project that terminated in January 2015 to help clarify the costs involved in data curation. Its goals were:

*"4C will help organisations across Europe to invest more effectively in digital curation and preservation. Research in digital preservation and curation has tended to emphasize the cost and complexity of the task in hand. 4C reminds us that the point of this investment is to realise a benefit, so our research must encompass related concepts such as 'risk', 'value', 'quality' and 'sustainability'."*

Its roadmap document[24] contains the following recommendations:

1. *Identify the value of digital assets and make choices;*
2. *Demand and choose more efficient systems;*
3. *Develop scalable services and infrastructure;*
4. *Design digital curation as a sustainable service;*

[24] See http://4cproject.eu/.

5. *Make funding dependent on costing digital assets across the whole lifecycle;*
6. *Be collaborative and transparent to drive down costs.*

With its leadership in providing scalable, sustainable services, HEP is well positioned to make key contributions in many of these areas. However, we must be aware of and plan for recommendation 5, which could have significant funding implications!

The Policy RECommendations for Open Access to Research Data in Europe (**RECODE**) project:

*"will leverage existing networks, communities and projects to address challenges within the open access and data dissemination and preservation sector and produce policy recommendations for open access to research data based on existing good practice."*

As for 4C, this was also an FP7-funded project that recently terminated, again with a final set of policy recommendations.

As has happened with publications, the most likely course of events is that the Open Access to data movement will gain momentum. However, given the above-mentioned LHC policies and the volumes of data involved, we need to be prepared to answer the following questions:

1. Is it financially affordable?
2. Is it technically implementable?
3. Is it scientifically (or educationally, or culturally) meaningful?

The answers to these questions may well vary with time and also depend on the implementation(s) that we choose: Open Access is just one step in the progression towards Open Data and finally "Open Knowledge".[25]

[25] An early but public draft of the Horizon 2020 2016-17 work programme states *"Research Infrastructures such as the ones on the ESFRI roadmap and others, are characterized by the very significant data volumes they generate and handle. These data are of interest to thousands of researchers across scientific disciplines and to other potential users via Open Access policies. Effective data preservation and open access for immediate and future sharing and re-use is a fundamental component of today's research infrastructures and Horizon 2020 actions."*

## PERICLES project

Pericles – for **P**romoting and **E**nhancing **R**euse of **I**nformation throughout the **C**ontent **L**ifecycle taking account of **E**volving **S**emantics – aims to address the challenge of ensuring that digital content remains accessible in an environment that is subject to continual change. Given the typically long lifetime of HEP experiments, adapting to such change is an inherent feature of any projects / experiments.

### Conferences & Workshops

Aside from project-oriented conferences, there are a number of conference series that are of relevance to DPHEP. These include:

- The annual International Digital Curation Conference (IDCC[26]);
- The annual International Conference on Digital Preservation (iPRES[27]);
- The biennial conference on "Ensuring Long-Term Data **P**reservation, and Adding **V**alue to Scientific and Technical Data" (PV[28]).

DPHEP has had a presence at both IDCC and PV conferences – the latter the most oriented towards practitioners and case studies. However, iPRES 2016 will be held in Bern and hence a presentation on DPHEP activities is foreseen.

Pointers to these and other relevant events are maintained in the DPHEP Indico pages: https://indico.cern.ch/category/4458/.

# Certification of Digital Repositories

Increasingly, the terms "trusted" or "certified" repositories are used: by data preservation projects, by communities requiring preservation services as well as by funding agencies in calls for project proposals. A number of methodologies exist – such as those from the Data Archiving and Networked Services (DANS) in the Netherlands, CODATA and finally a set of closely related ISO standards – that are in the process of being harmonised in the context of the RDA.

Following discussions at the WLCG Overview Board and following interest from the preservation community, a course was organised at CERN covering the ISO standards in this area, given by the authors of the standards involved.

There are three important ISO standards:

- **ISO 14721:2012** (OAIS – a reference model for what is required for an archive to provide long-term preservation of digital information)
- **ISO 16363:2013** (Audit and certification of trustworthy digital repositories – sets out comprehensive metrics for what an archive must do, based on OAIS)

[26] See http://www.dcc.ac.uk/events/international-digital-curation-conference-idcc.
[27] See http://www.ipres-conference.org/.
[28] See http://www.eumetsat.int/website/home/News/ConferencesandEvents/DAT_2447480.html.

- **ISO 16919:2014** (Requirements for bodies providing audit and certification of candidate trustworthy digital repositories – specifies the competencies and requirements on auditing bodies)

These three standards form a closely related family and an understanding of their principles and use will become increasingly important in establishing an internationally recognized set of trustworthy digital repositories.

Personnel followed this course from the WLCG Tier0 (CERN) and several WLCG Tier1 sites.

A checklist is available and it is foreseen – following further discussion in the WLCG and DPHEP communities – to proceed at least with a self-certification in 2016. This would help ensure that all of the necessary processes were in place, as well as identifying any gaps, for long-term preservation and re-use of HEP data. "Self-certification" discussions will form part of the DPHEP workshop that will be co-located with a WLCG workshop in Lisbon in February 2016.

Some of the metrics involved in obtaining certification are listed below.

| | Metric | Supporting Text | Examples |
|---|---|---|---|
| **3.1.1** | **THE REPOSITORY SHALL HAVE A MISSION STATEMENT THAT REFLECTS A COMMITMENT TO THE PRESERVATION OF, LONG TERM RETENTION OF, MANAGEMENT OF, AND ACCESS TO DIGITAL INFORMATION.** | This is necessary in order to ensure commitment to preservation and access at the repository's highest administrative level. | Mission statement or charter of the repository or its parent organization that specifically addresses or implicitly calls for the preservation of information and/or other resources under its purview; a legal, statutory, or government regulatory mandate applicable to the repository that specifically addresses or implicitly requires the preservation of information and/or other resources under its purview. |
| **3.1.3** | **THE REPOSITORY SHALL HAVE A COLLECTION POLICY OR OTHER DOCUMENT THAT SPECIFIES THE TYPE OF INFORMATION IT WILL PRESERVE, RETAIN, MANAGE AND PROVIDE ACCESS TO.** | This is necessary in order that the repository has guidance on acquisition of digital content it will preserve, retain, manage and provide access to. | Collection policy and supporting documents, Preservation Policy, mission, goals and vision of the repository. |
| **3.2.1** | **THE REPOSITORY SHALL HAVE IDENTIFIED AND ESTABLISHED THE DUTIES THAT IT NEEDS TO PERFORM AND SHALL HAVE APPOINTED STAFF WITH ADEQUATE SKILLS AND EXPERIENCE TO FULFIL THESE DUTIES.** | Staffing of the repository should be by personnel with the required training and skills to carry out the activities of the repository. The repository should be able to document through development plans, organizational charts, job descriptions, and related policies and procedures that the repository is defining and maintaining the skills and roles that are required for the sustained operation of the repository. | Organizational charts; definitions of roles and responsibilities; comparison of staffing levels to industry benchmarks and standards. |
| **3.3.1** | **THE REPOSITORY SHALL HAVE DEFINED ITS DESIGNATED COMMUNITY AND ASSOCIATED KNOWLEDGE BASE(S) AND SHALL HAVE THESE DEFINITIONS APPROPRIATELY ACCESSIBLE.** | This is necessary in order that it is possible to test that the repository meets the needs of its Designated Community. | A written definition of the Designated Community. |
| **3.3.2** | **THE REPOSITORY SHALL HAVE PRESERVATION POLICIES IN PLACE TO ENSURE ITS PRESERVATION STRATEGIC PLAN WILL BE MET.** | This is necessary in order to ensure that the repository can fulfill that part of its mission related to preservation | Preservation Policies; Repository Mission Statement. |
| **4.1.1** | **THE REPOSITORY SHALL IDENTIFY THE CONTENT INFORMATION AND THE INFORMATION PROPERTIES THAT THE REPOSITORY WILL PRESERVE.** | This is necessary in order to make it clear to funders, depositors and users what responsibilities the repository is taking on and what aspects are excluded. It is also a necessary step in defining the information which is needed from the information producers or depositors. | Mission statement; submission agreements/deposit agreements/deeds of gift; workflow and Preservation Policy documents, including written definition of properties as agreed in the deposit agreement/deed of gift; written processing procedures; documentation of properties to be preserved. |
| **4.3.4** | **THE REPOSITORY SHALL PROVIDE EVIDENCE OF THE EFFECTIVENESS OF ITS PRESERVATION ACTIVITIES.** | This is necessary in order to assure the Designated Community that the repository will be able to make the information available and usable over the mid-to-long-term. | Collection of appropriate preservation metadata; proof of usability of randomly selected digital objects held within the system; demonstrable track record for retaining usable digital objects over time; Designated Community polls. |
| **5.1.1** | **THE REPOSITORY SHALL IDENTIFY AND MANAGE THE RISKS TO ITS PRESERVATION OPERATIONS AND GOALS ASSOCIATED WITH SYSTEM INFRASTRUCTURE.** | This is necessary to ensure a secure and trustworthy infrastructure. | Infrastructure inventory of system components; periodic technology assessments; estimates of system component lifetime; export of authentic records to an independent system; use of strongly community supported software .e.g., Apache, iRODS, Fedora); re-creation of archives from backups. |
| **5.2.1** | **THE REPOSITORY SHALL MAINTAIN A SYSTEMATIC ANALYSIS OF SECURITY RISK FACTORS ASSOCIATED WITH DATA, SYSTEMS, PERSONNEL, AND PHYSICAL PLANT.** | This is necessary to ensure ongoing and uninterrupted service to the designated community. | Repository employs the codes of practice found in the ISO 27000 series of standards system control list; risk, threat, or control analysis. |

# Site / Experiment Status Reports (June 2015)

## Belle I & II

| Preservation Aspect | Status |
|---|---|
| **Bit Preservation** | Preamble: The central computing system at KEK is replaced every four years. The main user must be Belle II until the data taking ends (in 2024).<br>Belle : mDST (necessary for physics analysis) is stored on the disk as well as the tape library. The data is still frequently read by active analysis users. All data will be preserved by migrating to the next system. We experienced data loss in the previous data migration. Main causes of this trouble were the short migration period, miscommunication between researchers and operators and the lack of the validation scheme after the migration. We will improve the process of the future migration. |
| **Data** | Belle : raw data (1PB) and other format (incl. simulation, ~1.5PB) are stored at the KEK central computing system. This data will be migrated to the next system, at least (data will be preserved until 2020). However, there is no plan thereafter, because the data will be superseded by Belle II. And a full set of mDST was copied at PNNL in USA.<br>Belle II : data taking is not started yet. But raw data will be stored at KEK and another set will be copied in some places outside Japan. Also, the replicas of the mDST will be distributed to the world-wide collaborated computing sites. |
| **Documentation** | Belle : all documentation is stored in the local web server and INDICO system. They are still active and accessible, but not well catalogued at all.<br>Belle II : Using twiki, invenio, svn and INDICO system. |
| **Software** | Belle : software has been fixed since 2009 except for some patches. The baseline of the OS is still SL5, but it was migrated to SL6. In parallel, the Belle data I/O tool is developed and integrated in the Belle II software. Thanks to this, the Belle data can be analysed under the Belle II software environment. Other Belle handmade analysis tools are being integrated, too. Software version is maintained with SVN.<br>Belle II : basic features witch are necessary for the coming data taking have been implemented. But need more tuning and improvement. The software version is maintained by SVN. SL5/6 32/64-bits, Ubuntu 14.02 |
| | LTS are supported |
| **Uses Case(s)** | Continued analyses by Belle. |
| **Target Community(ies)** | Belle and Belle II |
| **Value** | Quantitative measures (# papers, PhDs etc) exist<br><br>Belle : During the data taking period (1999-2010), averaged number of journal publications is ~30 papers/year and the number of PhD is ~12/year.<br><br>After the data taking, a moderate decreasing tendency can be seen, but the analysis is still active. (~20 publications/year and ~7 PhDs/year). |
| **Uniqueness** | Belle : Comparing with the data from Hadron colliders, the Belle data has advantage to analyse the physics modes with missing energy and neutral particles. Until the Belle II starts, these data are unique as well as *BABAR*'s data.<br>Belle II : Belle data will be superseded by 2020. After that, the data must be unique samples. |
| **Resources** | Belle / Belle II : in some stage, the Belle data must be treated as a part of the Belle II data. And resources for the Belle data will be included in the Belle II computing/human resources. |
| **Status** | Construction of the Belle II detector/SuperKEKB accelerator as well as the Belle II distributed computing system. |
| **Issues** | A couple of items have to be implemented in the Belle II software framework to analyse the Belle data. Further check for the performance and reproducibility is also necessary. |
| **Outlook** | Expect to be able to analyse the Belle data within the Belle II software framework. It provides us the less human resource to maintain the Belle software, and the longer lifetime of the Belle data analysis. |

## BES III

| **Preservation Aspect** | **Status** |
|---|---|
| **Bit Preservation** | A MD5 integrity check is done when data is copied from disk to tape<br>Annual examination of tape library and LTO4 tapes (possibly moving to biennial due to risks to tapes) |
| **Data** | 2750TB acquired 2009-2014 with annual growth of 450TB leading to 3450TB in 2020.<br><br>Archive storage system based on CASTOR v1.8 with |
| | IBM3584 tape library, LTO 4<br><br>Current capacity for BESIII<br><br>• 2.7 PB, 2.2 PB used, 0.5 PB available<br><br>Remote replication of important raw data<br><br>• ~ 900 cartridges, 700 TB |
| **Documentation** | • DocDB: paper, technical notes, minutes…<br>• Hypernews: notifications of software release, paper publishing …<br>• Indico: Conference slides,<br>• Inspire: published paper |
| **Software** | BOSS is an integrated software package that includes all the blocks required in BESIII data processing.<br>For an old but stable version of BOSS, we preserve following items:<br><br>• A complete package of software,<br>• A runnable virtual machine image<br>• The puppet template and RPM repository from which a runnable OS is created,<br>• Release documents, book-keeping parameters…<br>• A functional validation is done according to the standard process of software release. |
| **Uses Case(s)** | |
| **Target Community(ies)** | |
| **Value** | |
| **Uniqueness** | |
| **Resources** | Since the experiment is still working, budget and FTEs are shared with the operation of computing centre |
| **Status** | |
| **Issues** | |
| **Outlook** | The experiment is expected to stop data taking at 2022 and Lifespan of preserved data is expected to be about 15 years after then. |

## HERA

| Preservation Aspect | Status |
|---|---|
| **Bit Preservation** | Well-defined data and MC sets mainly in plain ROOT format. Migration to new tape generations planned. |
| **Data** | Transferred to DPHEP area on DESY dCache. 2 tape copies (different media generations 1.2 PiB) plus disk cache (700 TiB) for on-going analyses. A copy of the ZEUS data is also available at MPCDF (Munich) and further copies are under consideration. |
| **Documentation** | Non-digital documentation catalogued and stored in the DESY library archive; some digitized. Collaboration software notes in INSPIREHEP. Important internal web pages collected to a web-server with static content. |
| **Software** | *"In the best of all worlds we would keep the software alive i.e. compilable on the latest Linux with the latest library versions"*<br>We now follow a "freezing approach", i.e. a VM with isolated storage and well defined set of external libs |
| **Uses Case(s)** | Continued analysis by members of the HERA collaborations |
| **Target Community(ies)** | Primarily the HERA collaborations, but also other physicists interested in HERA data |
| **Value** | Analyses, publications and PhDs continue to be produced |
| **Uniqueness** | Unique combination of initial state particles and energy |
| **Resources** | |
| **Status** | Transitioning from experiment-specific to institutional solutions |
| **Issues** | Webservers: tension between production needs and long-term archiving.<br>Do not underestimate the effort! Experiment expertise fades away quickly once funding stops.<br>**Data preservation must be prepared whilst effort is available!** |
| **Outlook** | Continued ability to analyse data until 2020 (when support for SL6 stops); Migration to SL7 could extend |
| | this; Tape archive will life on. |

## LEP

| Preservation Aspect | Status |
|---|---|
| **Bit Preservation** | "State of the art" bit preservation with regular scrubbing and migration to new media |
| **Data** | 2 copies on tape at CERN, an additional copy on disk (EOS) being setup.<br><br>Additional copies exist outside CERN (ALEPH, OPAL and partial copy for DELPHI) |
| **Documentation** | Being revisited – to be "archived" in CERN Document Server for long-term preservation |
| **Software** | To be published into CernVMFS |
| **Uses Case(s)** | Continued analyses by former collaboration members |
| **Target Community(ies)** | Primarily former collaboration |
| **Value** | Analyses, publications and PhDs continue to be produced |
| **Uniqueness** | Unique – until and unless certain FCC options are implemented |
| **Resources** | Minimal resources for "bit preservation" and storage |
| **Status** | |
| **Issues** | Dependency on CERNLIB (no longer maintained) |
| **Outlook** | Expect to be able to analyse data (ALEPH, DELPHI, OPAL) until at least 2020. Until 2030 should be possible with < (<) 1FTE / experiment / year |

| Preservation Aspect | Status (DELPHI) |
|---|---|
| **Bit Preservation** | State of the art bit preservation with regular scrubbing and migration to new media |
| **Data** | Two copies on tape at CERN, one copy on disk, and one copy at external institute (University of Cantabria, Santander) with tape archive. |
| **Documentation** | To be archived in CERN Document Server for long-term preservation. Some software documentation on CernVMFS alongside the code. |
| **Software** | Binaries and source code published into CernVMFS. Software CD on AFS at CERN. |
| **Uses Case(s)** | Continued analyses by former collaboration members. Discussing open access publication for education and outreach. |
| **Target Community(ies)** | Primarily former collaboration. Open access data targeted at physicists and students, to be approved by the collaboration. |
| **Value** | Analyses, publications and PhDs continue to be produced |
| **Uniqueness** | Unique – until and unless certain FCC options are implemented |
| **Resources** | Minimal resources for bit preservation and storage. WLCG infrastructure for software on CVMFS. |
| **Status** | Data and simulated events are available on EOS. A DELPHI collaborator using CernVM may reconstruct and analyse the data and generate Monte Carlo events. |
| **Issues** | Dependency on CERNLIB and 32-bit architecture (no longer maintained) |
| **Outlook** | Expect to be able to analyse data until at least 2020. Until 2030 should be possible with < (<) 1FTE / year |

## Tevatron

| **Preservation Aspect** | **Status** |
|---|---|
| **Bit Preservation** | All data migrated to T10k technology (2 ½ years).<br>Data integrity checks: After each copy during migration; Periodic reads from each tape.<br>Long term future preservation of CDF data at INFN-CNAF, developed in collaboration with CDF and FNAL SCD. |
| **Data** | Two copies of raw data at FNAL, in different locations. In case of damage/loss analysis ntuples can be reproduced and/or eventually recovered from CNAF. |
| **Documentation** | All online webpages and code archived, still accessible from CDF webpages. |
| **Software** | All online webpages and code archived, still accessible from CDF webpages.<br><br>At the time of Tevatron shutdown<br>• all code in frozen releases or in CVS repositories<br>• based on 32-bit frameworks built on Scientific Linux 5 (but with compatibility libraries to older OSs)<br>Long term future solution: build legacy release that contains no pre-SL6 libraries<br><br>CVMFS for code distribution |
| **Uses Case(s)** | Continued analyses by former collaboration members |
| **Target Community(ies)** | Primarily former collaboration |
| **Value** | Quantitative measures (# papers, PhDs etc.) exist |
| **Uniqueness** | Unique initial state vs. LHC; Multiple energy collisions (300, 900 and 1960 GeV) |
| **Resources** | FNAL R2DP project budgeted 4 (3) FTE in 2013, 3 (2.1) in 2014 and 0.3 (0.4) in 2015. (Expenditure) |
| **Status** | R2DP project complete |
| **Issues** | Both CDF and D0 use Oracle → licence cost is a long-term future challenge. Migration to open source db would require considerable human effort (need to rewrite the analysis software) |
| **Outlook** | Goal: Complete analysis capability (DPHEP "level 4") through Nov 2020 (SL6 EOL) and beyond. |

## *BABAR*

| Preservation Aspect | Status |
|---|---|
| **Bit Preservation** | 2.7PB of data of which 2PB (budget constraints) will be migrated to new media when supported by SLAC |
| **Data** | Data is stored on tape at SLAC and CC-IN2P3 (back-up only); Active data on disk accessed via xrootd. |
| **Documentation** | All the most used and fundamental information have |
| | been checked, updated and moved to a Media Wiki server, the *BABAR* WIKI. Internal documents are stored on disk and backed up on tape and are accessible to the Collaboration via web applications. All published papers are available through the *BABAR* web and arXiv/inSPIRE. |
| **Software** | Software releases: C++ Object Oriented 32 bit. They build and run on SL4, SL5, SL6 32 and 64 bit (or corresponding RH). Other software: Tcl, Perl, Phyton, SQL (MySQL, Oracle). Software releases, even in frozen virtual environment, still preserve their full capability to handle data processing, data analysis, and future extensions (new analyses, new physics models). |
| **Uses Case(s)** | Continued analyses by *BABAR* Collaboration members. |
| **Target Community(ies)** | Primarily the *BABAR* Collaboration. |
| **Value** | Quantitative measures (# papers, PhDs, etc.) exist.<br><br>More than 30 analyses are on track for publication, about 20 have less clear future. |
| **Uniqueness** | Data will not be superseded by LHC – some by Belle II (for example, not the Y(3S) data sample). |
| **Resources** | 0.35 FTE computing support for *BABAR* at SLAC by end 2015 |
| **Status** | The *BABAR* Collaboration is still very active and engaged even if resources are dwindling. |
| **Issues** | Much of the hardware is aging; Sun OS 5.10 support will stop at SLAC within 2 years and corresponding h/w will be decommissioned |
| **Outlook** | Aim to preserve data for on-going analyses until 2018 with extension to 2020+ to match Belle II schedule.<br><br>The technology at the base of the future operating model will be virtualization – all the services now running on physical hardware will soon run on virtual machines |

## LHC

| **Preservation Aspect** | **Status (Generic "WLCG")** |
|---|---|
| **Bit Preservation** | "State of the art" bit preservation with regular scrubbing |
| | and migration to new media |
| **Data** | Stored at WLCG Tier0 with additional copies across WLCG Tier1 sites |
| **Documentation** | |
| **Software** | "Published" into CernVMFS |
| **Uses Case(s)** | "Standard" |
| **Target Community(ies)** | Re-use of data within the collaboration(s), sharing with the wider scientific community, Open Access releases |
| **Value** | Landmark discoveries already made; significant potential for future "BSM" discoveries |
| **Uniqueness** | Unique data sets (both pp and HI) being acquired now - ~2035. Probably unique until "FCC" (2035-2050?) |
| **Resources** | Computing resources via Resource Review Board |
| **Status** | |
| **Issues** | Effort within the experiments is hard to find |
| **Outlook** | On-going activity on analysis capture and reproducibility. Regular public releases (according to individual experiment policies) and "master classes" |

| **Preservation Aspect** | **Status (ALICE)** |
|---|---|
| **Bit Preservation** | On tape: data integrity check during each access request<br><br>On disk: periodically integrity checks |
| **Data** | 7.2 PB of raw data were acquired between 2010 and 2013 which is stored on tape and disk in 2 replicas. |
| **Documentation** | ALICE analysis train system & bookkeeping in Monalisa DB: for the last 3-4 years<br><br>Short introduction along with the analysis tools on Opendata |
| **Software** | The software package "AliRoot" is published on CVMFS<br><br>For the Open Access the data and code packages are available on Opendata (http://opendata.cern.ch/) |
| **Uses Case(s)** | Educational purposes like the CERN Master Classes<br><br>Outreach activities |
| **Target Community(ies)** | Re-use of data within the collaboration(s), sharing with the wider scientific community, Open Access releases |
| **Value** | Analysis, publications and PhDs continue to be produced |
| **Uniqueness** | Unique data sets from the LHC in pp and HI<br><br>Similar data can only be collected by the other LHC experiments |
| **Resources** | Since the experiment is still working, budget and FTEs are shared with the operation of computing centre |
| **Status** | First data from 2010 has been released to the public<br><br>(8 TB ≈ 10% of data)<br><br>Some analysis tools are available on Opendata for the CERN Master class program |
| **Issues** | Improve user interface<br><br>The interaction with the open-access portal is very slow due to long communication times. E.g. the uploading of data is done by some people in the IT department. The interaction via an automated website would be faster. |
| **Outlook** | Ongoing analysis within the collaboration<br><br>Making realistic analysis available on the open-access portal<br><br>Deployment of more data |

| Preservation Aspect | Status (ATLAS) |
|---|---|
| **Bit Preservation** | Non-Reproducible data exist in two or more geographically disparate copies across the WLCG. The site bit preservation commitments are defined in the WLCG Memorandum of Understanding[29]. All data to be reprocessed with most recent software to ensure longevity. |
| **Data** | Non-reproducible: RAW physics data, calibration, metadata, documentation and transformations (jobs). Derived data: formats for physics analysis in |

[29] WLCG MOU: http://wlcg.web.cern.ch/collaboration/mou

| | |
|---|---|
| | collaboration, formats distributed for education and outreach. Greatly improved by common derived data production framework in run 2. Published results in journals and HEPDATA. Sometimes with analysis published in Rivet and RECAST. Format lifetimes are hard to predict, but on current experience are 5-10 years, and changes are likely to coincide with the gaps between major running periods. |
| **Documentation** | Software provenance of derived data stored in AMI database. Numerous twikis available describing central and analysis level software. Interfaces such as AMI and COMA contain metadata.<br>The publications themselves are produced via the physics result approval procedures set out in ATL-GEN-INT-2015-001 held in CDS; this sets out in detail the expected documentation within papers and the supporting documentation required. |
| **Software** | Compiled libraries and executable of the "Athena" framework are published on CVMFS. Software versioning is maintained on the CERN subversion server. |
| **Uses Case(s)** | Main usage of data: future analysis within the collaboration<br>Further usage: review in collaboration and potential for outreach |
| **Target Community(ies)** | Re-use of data (new analyses) within the collaboration, open access sharing of curated data |
| **Value** | Publications by the collaboration. Training of PhDs |
| **Uniqueness** | Unique data sets (both pp and HI) being acquired between now and 2035. Similar data only acquired by other LHC experiments |
| **Resources** | The active collaboration shares the operational costs with the WLCG computing centres. |
| **Status** | ATLAS replicates the non-reproducible data across the WLCG and maintains database of software provenance to reproduce derived data. Plans to bring run 1 data to run 2 status. Master-classes exercises available on CERN Open Data Portal, expansion considered. Some analyses published on Rivet/RECAST. |
| **Issues** | Person-power within the experiment is hard to find. Validation of future software releases against former processing crucial. No current plans beyond the lifetime of the experiment. |
| **Outlook** | On-going development of RECAST with Rivet and collaboration with CERN IT and the other LHC experiments via the CERN Analysis Portal as solution to problem of analysis preservation. |

| Preservation Aspect | Status (CMS) |
|---|---|
| **Bit Preservation** | Follow WLCG procedures and practices<br><br>Check checksum in any file transfer |
| **Data** | RAW data stored at two different T0<br>• 0.35 PB 2010<br>• 0.56 PB 2011<br>• 2.2 PB 2012<br>• 0.8 PB heavy-ion 2010-2013<br><br>Legacy reconstructed data (AOD):<br>• 60 TB 2010 data reprocessed in 2011 with CMSSW42 (no corresponding MC)<br>• 200 TB 2011 and 800 TB 2012 reprocessed in 2013 with CMSSW53 (with partial corresponding MC for 2011, and full MC for 2012)<br><br>Several reconstruction reprocessings<br><br>The current plan: keep a complete AOD reprocessing (in addition to 2×RAW)<br>• no reconstructed collision data have yet been deleted, but deletion campaigns are planned.<br>• most Run 2 analyses will use miniAOD's which are significantly smaller in size<br><br>Open data: 28 TB of 2010 collision data released in 2014, and 130 TB of 2011 collision data to be released in 2015 available in CERN Open Data Portal (CODP)<br><br>Further public releases will follow. |
| **Documentation** | Data provenance included in data files and further information collected in CMS Data Aggregation System (DAS)<br><br>Analysis approval procedure followed in CADI<br><br>Notes and drafts stored in CDS<br><br>Presentations in indico |
| | User documentation in twiki serves mainly the current operation and usage<br><br>Basic documentation and examples provided for open data users in CODP<br><br>Set of benchmark analyses reproducing published results with open data in preparation, to be added to CODP |
| **Software** | CMSSW open source and available in github and in CVFMS<br><br>Open data: VM image (CERNVM), which builds the appropriate environment from CVFMS, available in COPD |
| **Uses Case(s)** | Main usage: analysis within the collaboration<br><br>Open data: education, outreach, analysis by external users |
| **Target Community(ies)** | Main target: collaboration members<br><br>Open data: easy access to old data for collaboration members and external users |
| **Value** | Data-taking and analysis is on-going, more than 400 publications by CMS<br><br>Open data: educational and scientific value, societal impact |
| **Uniqueness** | Unique, only LHC can provide such data in any foreseeable time-scale |
| **Resources** | Storage within the current computing resources<br><br>Open data: storage for the 2010-2011 open data provided by CERN IT, further requests to be allocated through RRB |
| **Status** | Bit preservation guaranteed in medium term within the CMS computing model and agreements with computing tiers, but the long-term preservation beyond the life-time of the experiment not yet addressed (storage, agreements, responsibilities),<br><br>Open data release has resulted in<br><br>• data and software access independent from the experiment specific resources |
| | • a timely capture of the basic documentation, which, although limited and incomplete, makes data reuse in long term possible<br><br>common solutions and services. |
| **Issues** | Competing with already scarce resources needed by an active experiment.<br><br>Knowledge preservation, lack of persistent information of the intermediate analysis steps to be addressed by the CERN Analysis Preservation framework (CAP)<br><br>• CMS has provided input for the data model and user interface design, and defining pipelines for automated ingestion from CMS services.<br>• The CAP use-cases are well acknowledged by CMS.<br>• CAP will be valuable tool to start data preservation while the analysis is active.<br><br>Long-term reusability: freezing environment (VM) vs evolving data: both approaches will be followed and CMS tries to address the complexity of the CMS data format |
| **Outlook** | Impact of the open data release very positive<br><br>• well received by the public and the funding agencies<br>• no unexpected additional workload to the collaboration<br>• the data are in use.<br><br>Excellent collaboration with CERN services developing data preservation and open access services and with DASPOS<br><br>• Common projects are essential for long-term preservation<br>• Benefit from expertise in digital archiving and library services<br>• Fruitful discussion with other experiments.<br><br>Long-term vision and planning is difficult for ongoing experiments: |
| | • DPHEP offers a unique viewpoint.<br><br>Next steps for CMS:<br><br>• stress-test CERN Open Data Portal with the new data release<br><br>develop and deploy the CMS-specific interface to CERN Analysis Preservation framework |

| Preservation aspect | Status (LHCb) |
|---|---|
| **Bit preservation** | Data and MC samples are stored on tape and on disk. Two copies of raw data on tape ; 1 copy on tape of full reconstructed data (FULL.DST, which contains also raw data) ;4 copies of stripped data (DST) on disk for the last (N) reprocessing. Two copies for the N-1 reprocessing. One archive replica on tape. |
| **Data** | For the long term future, LHCb plans to preserve only a legacy version of data and MC samples. Run 1 legacy data: 1.5 PB (raw), 4 PB FULL.DST, 1.5 stripped DST. Run 1 legacy MC : 0.8 PB DST.<br><br>Open data: LHCb plans to make 50% of analysis level data (DST) public after 5 years, 100% public 10 years after it was taken. The data will be made public via the Open Data portal (http://opendata.cern.ch/)<br><br>Samples for educational purposes are already public for the International Masterclass Program and accessible also via the Open Data portal (For Education area). |
| **Documentation** | Data: dedicated webpages for data and MC samples, with details about all processing steps.<br><br>Software : twiki pages with software tutorials, mailing-lists.<br><br>Documentation to access and analyse masterclasses samples is available on LHCb webpage and on the OpenData portal. |
| **Software** | Software is organised as hierarchy of projects containing packages, each of which contains some c++ or python code. Three projects for the framework (Gaudi, LHCb, Phys), several "component" projects for algorithms (e.g. Lbcom, Rec, Hlt, Analysis), one project per application containing the application configuration (e.g. Brunel, Moore, DaVinci).<br>Software repository: SVN.<br>Open access: once data will be made public, software to work with DST samples will be released with the necessary documentation.<br>A virtual machine image of LHCb computing environment allows to access and analyse the public samples available on the Open Data portal |
| **Use cases** | New analysis on legacy data ; anaysis reproduction ; outreach and education. |
| **Targeted communities** | LHCb collaboration ; physicists outside the collaboration ; general public. |
| **Value** | LHCb complementary to other LHC experiments. |
| **Uniqueness** | Unique samples of pp an HI collisions collected in the forward region. |
| **Resources** | Dedicated working group within LHCb computing group. |
| **Status** | Legacy software and data releases defined. Development of a long-term future validation framework ongoing. Masterclasses samples and analysis software available via the Open Data portal. Collaboration with CERN IT and other LHC experiments for the development of an analysis preservation framework. |
| **Issues** | Main issue is manpower. |
| **Outlook** | Collaboration with CERN IT and other LHC experiments on the Open Data portal and the analysis preservation framewok.<br><br>Enrich the Open Data portal with additional masterclass exercise and real LHCb analysis. Exploit VM technology to distribute LHC computing environment. |

# Towards a Data Preservation Strategy for CERN Experiments

The updated Strategy for European Particle Physics[30], approved by Council in May 2014, states that *"infrastructures for … data preservation … should be maintained and further developed."*

In order to implement this strategy, the following proposals are currently under discussion. (The numbering reflects the draft proposal, where the paragraph above is point 1.):

2. Such infrastructures include *digital repositories*, where *copies* or *replicas* of the data are kept.
3. As host laboratory, it is expected that (from now on?) a copy of all data acquired by CERN experiments *and* targeted for long-term preservation be stored in the CERN digital repository. This will typically include all raw data and the final reprocessing pass and associated Monte Carlo datasets.

[30] See http://council.web.cern.ch/council/en/EuropeanStrategy/ESParticlePhysics.html.

4. It is strongly recommended that one or more copies of the above data are maintained outside, at or spread over institutes that form part of the collaboration.
5. In order to ensure sufficient reliability and adherence to "best practices", it is recommended that such repositories follow agreed guidelines / standards – this is currently being discussed in the context of WLCG for LHC data.
6. These guidelines not only include policies for the management of the repository itself, but also on access to data in the repository (adherence to agreed access policies and terms of use), as well as the *ingest* process, when data is "entered" into the repository. The latter is to ensure that appropriate and supported data formats are used, there is sufficient documentation, meta-data and other materials to permit use by the designated communities, and so forth.
7. The above recommendations could become part of a default strategy for CERN experiments, with implementation details – including variances on the above – provided in the Data Management Plan (DMP) for that experiment. DMPs are increasingly required by funding agencies for new and/or repeat funding and can be expected to be quasi-mandatory in the future.
8. As a minimum, the DMP of an experiment should detail the policy for storing replicas of data and the recovery mechanisms, both during and after the active lifetime of the associated collaboration.
9. These basic recommendations are expected to be supplemented by others – e.g. on "knowledge capture and preservation" – as we gain experience with preserved and open access data.

It is foreseen that this proposal will be discussed at CERN's scientific committees – most likely starting with the LHCC, as an implementation based on the WLCG Tier0 and Tier1 sites could be a reality in the short to medium term.

## Changes With Respect to the Blueprint

With respect to the DPHEP Blueprint, the following observations can be made:

- The pervasive use of INSPIREHEP and other Invenio-based solutions will come as no surprise;
- "Bit Preservation" (and loss) is more clearly defined, with extensive practical experience, albeit different implementations due to site preferences and requirements (hardware choices, funding schemes etc.);
- Virtualisation is more prominent with a better defined timeline (circa ten years);
- The use of CVMFS is a clear success story;
- Cost models and business cases are better understood, with quantitative measures across a variety of experiments;
- "Open Access" policies, embargo periods and the like are new but match well with the *"Zeitgeist"*;
- A variety of "end-of-life" scenarios have been realised: moving from experiment to site support, from host institution to former collaboration members and even porting to new systems and services, such as EUDAT.

These developments, as well as the concrete experience over the past three years, positions the DPHEP Collaboration well to make clear recommendations to future projects and experiments.

## Lessons for Future Circular Colliders / Experiments

**The main message – from Past and Present Circular Colliders to Future ones – is that it is never early to consider data preservation: early planning is likely to result in cost savings that may be significant. Furthermore, resources (and budget) beyond the data-taking lifetime of the projects must be foreseen from the beginning.**

Beyond that, the activities of numerous data preservation activities worldwide can be used as a guide to the type of activities, services and support that is required.

In other words, at least "observer status" from the FCC activities in the DPHEP Collaboration is to be strongly recommended.

For other future and / or current experiments the recommendations are similar:

- Align yourselves with the overall strategy and even implementation of other data preservation activities at your institute / laboratory or globally;
- Adopt mainstream and supported technologies where-ever possible;
- Understand the target communities for your data preservation activities, the Use Cases and the expect benefits and outcomes;
- Try to understand the costs – in particular those that are specific to your collaboration (and not "external" – e.g. host laboratory bit preservation services);

- Data preservation services and support for the LHC experiments can be expected to be provided for several decades: this may be a good place to start.

## Future Activities

Over the next period, one can expect progress to be made in the following areas:

- The establishment of a formal policy regarding data preservation for CERN experiments (perhaps linked to the approval process through the Research Board);
- At least a "self-audit" for the CERN Tier0 and WLCG Tier1 sites in the context of the WLCG project;
- Further developments in terms of Analysis Capture and Preservation;
- Further releases of Open Data through the CERN Open Data Portal;
- Harmonization of similar activities across various laboratories and projects;
- Extension of DPHEP's activities to consider also those of potential FCCs;
- Clarifications regarding funding – of particular importance to past experiments where resources have already become sub-optimal;
- The continuation of regular meetings and workshops, aligning as much as possible with related events (WLCG, CHEP, HEP Software Foundation etc.);
- Further input to the next round of ESPP – building on concrete experience, results and remaining challenges.

The long-term management of the Collaboration also has to be considered – up to 2020 but also beyond.

## Outlook and Conclusions

There are clearly many similarities in the approaches being taken, the technologies deployed and the issues encountered. Regular reporting of results (possibly synchronised with major events such as CHEP) should be sufficient to ensure that coordinated approaches remain and that duplication is minimised.

The following quote[31] is traditionally attributed to Leslie Lamport – the initial author of LaTeX and an expert on distributed computing systems.

> A distributed system is one in which the failure of a computer you didn't even know existed can render your own computer unusable.

This reminds us that data preservation is inherently unstable – with many components and dependencies, constant attention is required to ensure that the entire "system" remains usable. Some changes may be relatively minor, such as a name change in a webserver. Others can be much more disruptive, such as major change in operating system (think VAX/VMS to Unix) or programming language – even a standard-conforming language changes over time, with some constructs being first deprecated, then obsolete and finally unsupported.

[31] See http://research.microsoft.com/en-us/um/people/lamport/pubs/distributed-system.txt.

Given the cost of today's storage and the likely evolution, there is no inherent cost why "data" cannot be stored more or less indefinitely. What is harder is to capture the necessary knowledge and validation procedures so that it can be used over long periods of time.

The "natural periodicity" of recent collider generations – some twenty years – is perhaps all one can hope for in terms of affordable data preservation. (Most LEP data – that of ALEPH, DELPHI and OPAL – may be usable somewhat longer, perhaps up to 25 / 30 years). Beyond that, re-use of the data will probably still be possible but may require a larger investment to "resuscitate", as has been done on rare (one?) occasion(s), notably for the JADE[32] experiment at the PETRA storage ring in DESY.

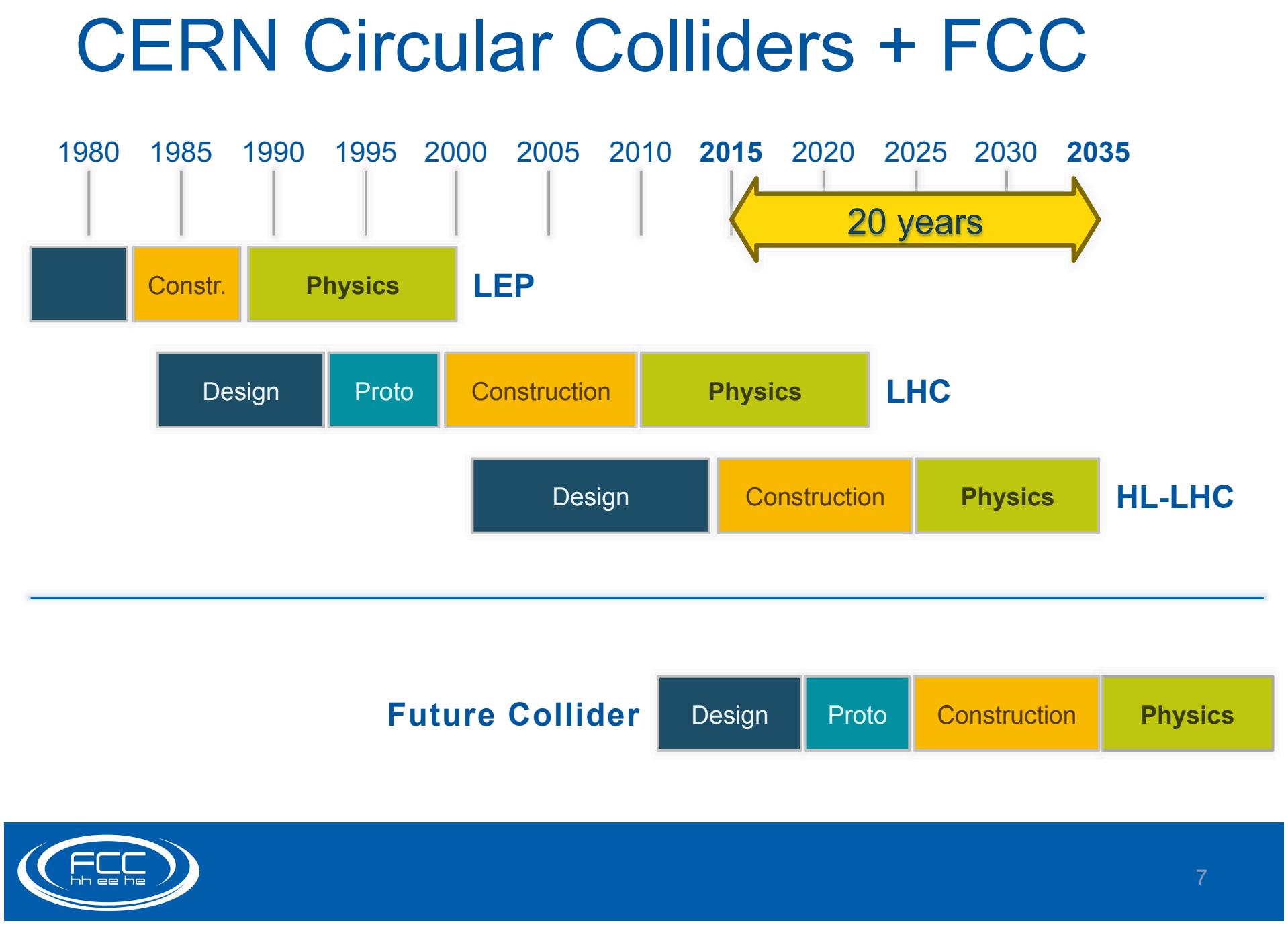

Figure 20 - Timeline of Major Colliders at CERN (+ "FCC")

[32] See https://wwwjade.mpp.mpg.de/ and the DPHEP Blueprint for further information.

# Appendix A – The DPHEP Collaboration

| **DPHEP Partner**<br>**(May 2014 unless specified)** | **Location** | **Contact person** |
|---|---|---|
| European Organization for Nuclear Research, **CERN** | Switzerland | J. Shiers |
| Deutsches Elektronen-Synchrotron, **DESY** | Germany | D. South |
| Helsinki Institute of Physics, **HIP** | Finland | K. Lassila-Perini |
| Institute of High Energy Physics, **IHEP** | China | G. Chen |
| Institut national de physique nucléaire et de physique des particules, **IN2P3** | France | G. Lamanna |
| Institute of Particle and Nuclear Studies, High Energy Accelerator Research Organisation, **IPNS, KEK** | Japan | T. Hara |
| Max Planck Institut für Physik, **MPP** | Germany | S. Kluth |
| Institute of Particle Physics, **IPP**<br>**(June 2015)** | Canada | R. Sobie |
| Science and Technology Facilities Council, **STFC**<br>**(July 2015 – pending CB approval)** | UK | J. Bicarregui |
| Istituto Nazionale di Fisica Nucleare, **INFN**<br>**(pending signature)** | Italy | M. Maggi |

US labs might sign a "Letter of Intent" apparently? (Although they did sign the WLCG MoU).

# Appendix B – The DPHEP Implementation Board

(CERN e-group DPHEP-IB)

Alicia Calderon Tazon <Alicia.Calderon@cern.ch> Self added member

Andrew Branson <andrew.branson@cern.ch>

Andrii Verbytskyi <andrii.verbytskyi@cern.ch> Self added member

Benedikt Hegner <Benedikt.Hegner@cern.ch>

<boj@fnal.gov>

Concetta Cartaro <cartaro@slac.stanford.edu>

<charles.f.vardeman.1@nd.edu>

David Colling <d.colling@imperial.ac.uk>

David Michael South <david.south@cern.ch>

<david.south@desy.de>

<denisov@to.infn.it>

Cristinel Diaconu <diaconu@cppm.in2p3.fr>

<dich@mail.desy.de>

<diesburg@fnal.gov>

Dirk Krucker <dirk.krucker@cern.ch> Self added member

<dirk.kruecker@desy.de>

Frank Berghaus <frank.berghaus@cern.ch>

<frank.berghaus@gmail.com>

<gang.chen@ihep.ac.cn>

<genevieve.romier@idgrilles.fr>

Gerardo Ganis <Gerardo.Ganis@cern.ch>

Gerhard Mallot <Gerhard.Mallot@cern.ch>

German Cancio Melia <German.Cancio.Melia@cern.ch>

<homer@slac.stanford.edu>

Jakob Blomer <Jakob.Blomer@cern.ch> Self added member

Jamie Shiers <Jamie.Shiers@cern.ch>

<jareknabrzyski@gmail.com>

Jetendr Shamdasani <Jetendr.Shamdasani@cern.ch> UWE

John Harvey <John.Harvey@cern.ch> Self added member

Kati Lassila-Perini <Katri.Lassila-Perini@cern.ch>

<kherner@fnal.gov>

<m.wing@ucl.ac.uk>

<marcello.maggi@ba.infn.it>

Marcello Maggi <Marcello.Maggi@cern.ch>

Marco Cattaneo <Marco.Cattaneo@cern.ch>

Maria Girone <Maria.Girone@cern.ch>

<matthew.viljoen@stfc.ac.uk>

Matthias Schroeder <Matthias.Schroder@cern.ch>

<meenakshi_narain@brown.edu>

<michael.d.hildreth.2@nd.edu>

Mihaela Gheata <Mihaela.Gheata@cern.ch>

Miika Tuisku <miika.tuisku@iki.fi>

Patricia Sigrid Herterich <patricia.herterich@cern.ch>

<Pere.Mato@cern.ch>

Peter Clarke <peter.clarke@ed.ac.uk>

Predrag Buncic <Predrag.Buncic@cern.ch>

Richard Mcclatchey <Richard.Mcclatchey@cern.ch>

<Roger.Jones@cern.ch>

Salvatore Mele <Salvatore.Mele@cern.ch>

<silvia.amerio@pd.infn.it>

<southd@mail.desy.de>

Sunje Dallmeier-Tiessen <sunje.dallmeier-tiessen@cern.ch>

<takanori.hara@kek.jp>

Tibor Simko <Tibor.Simko@cern.ch>

<Tim.Smith@cern.ch>

<tpmccauley@gmail.com>

Ulrich Schwickerath <Ulrich.Schwickerath@cern.ch>

<wolbers@fnal.gov>

<yves.kemp@desy.de>